\documentclass[10pt,letterpaper,onecolumn,oneside]{article}
\usepackage[top=0.85in,left=1.75in,footskip=0.75in,marginparwidth=2in]{geometry}

\usepackage{amsmath}
\usepackage{amssymb}
\usepackage{yhmath}
\usepackage[english]{babel}
\usepackage{graphicx}
\usepackage{epstopdf, epsfig}
\usepackage{comment}
\usepackage{color}
\usepackage{setspace}
\usepackage{siunitx}
\usepackage{booktabs}
\usepackage{units}
\usepackage[normalem]{ulem}
\usepackage{comment}
\usepackage{subfig}
\usepackage[makeroom]{cancel}
\usepackage{xspace}
\usepackage{float}
\usepackage{adjustbox}

\newcommand{\bs}{\boldsymbol}  

\DeclareSIUnit\Molar{\textsc{m}} 

\newcommand{\fd}[3]{\frac{\mathrm{d}^{#3}#1}{\mathrm{d}#2^{#3}}} 
\newcommand{\fp}[3]{\frac{\partial^{#3}#1}{\partial#2^{#3}}} 
\newcommand{\gf}{\phi} 

\newcommand{\Peclet}{P\'eclet\xspace}
\newcommand{\Leveque}{L\'ev\^eque\xspace} 
\newcommand{\arccosh}{\mathrm{arccosh}}

\newcommand{\paolo}[1]{\textcolor{black}{#1}}
\newcommand{\julien}[1]{\textcolor{black}{#1}}
\newcommand{\francois}[1]{\textcolor{black}{#1}}
\newcommand{\fernando}[1]{\textcolor{black}{#1}}

\newcommand{\eg}{{e.g.}\xspace}
\newcommand{\ie}{{i.e.}\xspace}
\newcommand{\fig}{figure\xspace}
\newcommand{\App}{appendix\xspace}

\usepackage[utf8]{inputenc}

\usepackage[colorlinks]{hyperref}

\usepackage[right]{lineno}

\usepackage{microtype}
\DisableLigatures[f]{encoding = *, family = * }

\usepackage{changepage}

\usepackage[aboveskip=1pt,labelfont=bf,labelsep=period,singlelinecheck=off,font=small]{caption}

\makeatletter
\renewcommand{\@biblabel}[1]{\quad#1.}
\makeatother

\usepackage{lastpage,fancyhdr,graphicx}
\definecolor{Gray}{gray}{.25}

\usepackage{sidecap}

\usepackage{wrapfig}
\usepackage[pscoord]{eso-pic}
\usepackage[fulladjust]{marginnote}
\reversemarginpar

\usepackage{textcomp}

\usepackage[round, numbers, sort&compress]{natbib}

\begin{document}
\thispagestyle{plain}

\vspace*{0.35in}

\begin{flushleft}
{\Large
\textbf\newline{A theory for the slip and drag of superhydrophobic surfaces with surfactant}
}
\newline
\\
{Julien R. Landel}\textsuperscript{a},
{Fran\c cois J. Peaudecerf}\textsuperscript{b,}\footnote{Present address: Institute of Environmental Engineering, Department of Civil, Environmental and Geomatic Engineering, ETH Z\"urich, 8093 Z\"urich, Switzerland},
{Fernando Temprano-Coleto}\textsuperscript{c},\\
{Fr\'ed\'eric Gibou}\textsuperscript{c},
{Raymond E. Goldstein}\textsuperscript{b},
{and Paolo Luzzatto-Fegiz}\textsuperscript{c,}\footnote{Email address: fegiz@engineering.ucsb.edu}
\\
\bigskip
\textsuperscript{a} Department of Mathematics, University of Manchester, \\Oxford Rd, Manchester M13 9PL, UK
\\
\textsuperscript{b} Department of Applied Mathematics and Theoretical Physics, \\Centre for Mathematical Sciences, University of Cambridge, \\Wilberforce Road, Cambridge CB3 0WA, UK
\\
\textsuperscript{c} Department of Mechanical Engineering, University of California, Santa Barbara, \\Santa Barbara CA 93106, USA
\\
\bigskip
\today

\end{flushleft}

\section*{Abstract}
Superhydrophobic surfaces (SHSs) have the potential to reduce drag at solid boundaries. However, multiple independent studies have recently shown that small amounts of surfactant, naturally present in the environment, can induce Marangoni forces that increase drag, at least in the laminar regime. To obtain accurate drag predictions, one must solve the mass, momentum, bulk surfactant and interfacial surfactant conservation equations. This requires expensive simulations, thus preventing surfactant from being widely considered in SHS studies. To address this issue, we propose a theory for steady, pressure-driven, laminar, two-dimensional flow in a periodic SHS channel with soluble surfactant. We linearise the coupling between flow and surfactant, under the assumption of small concentration, finding a scaling prediction for the local slip length. To obtain the drag reduction and interfacial shear, we find a series solution for the velocity field by assuming Stokes flow in the bulk and uniform interfacial shear. We find how the slip and drag depend on the \paolo{nine} dimensionless groups \paolo{that together characterize the} surfactant transport near SHSs, the gas fraction and the \paolo{normalized} interface length. Our model agrees with numerical simulations spanning orders of magnitude in each dimensionless group. The simulations also provide the constants in the scaling theory. Our model significantly improves predictions relative to a surfactant-free one, which can otherwise overestimate slip and underestimate drag by several orders of magnitude. Our slip length model can provide the boundary condition in other simulations, thereby accounting for surfactant effects without having to solve the full problem.

\vspace*{0.7in}

%


	
	
    
    
    
    

\section{Introduction}
Superhydrophobic surfaces (SHSs) consist of hydrophobic coatings equipped with micro- or nano-scale textures, such that a layer of air \citep[known as a ``plastron''; see \eg][]{Shirtcliffe2006-hi} is retained when the surface is submerged in water \citep[see \eg the reviews of][]{Quere_ARMR_2008,Rothstein:2010im,Samaha12,Lee2016-jn,Bhushan2018-jz}. The air layer is held in place by the texture, with the upper edges of the micro- or nano-structures making contact with the water. Since air is approximately 50 times less viscous than water, the plastron has often been approximated as a shear-free surface in analytical models~\citep{Philip_ZAMP_1972a,Philip_ZAMP_1972b,Lauga_Stone_JFM_2003,Cottin-Bizonne2004-ea,Ybert2007-mi,sbragaglia07,teo09,Davis2009-vp,Crowdy-2016uk}, leading to the expectation that SHSs could achieve very large drag reduction. Potential applications include high-Reynolds-number, turbulent flows \citep[\eg][]{Park2013-fj,Park2014-an,Ling2016-xj,Cartagena2018-gb,Gose2018-wo,Rastegari2018-an}, as well as low-Reynolds-number, internal flows, which are the focus of the present paper \citep[\eg][]{Lauga_Stone_JFM_2003,Ou2004-vk,Ou2005-ph,Ybert2007-mi,Kim2012-iw,Bolognesi2014-vw,Schaffel2016-mh,peaudecerf17,Song2018-uw}. 
At low Reynolds numbers, the use of SHSs has been proposed to reduce what are otherwise very large pressure differences across microchannels, as is the case in microfluidic devices or in micro-cooling applications \citep{cheng15,lam15,kirk17}, as well as to minimize Taylor dispersion in the chemical or biological analysis of species \citep{Cottin-Bizonne2004-ea}.

However, laminar-flow experimental results have been mixed. While early works reported large drag reduction~\citep[\eg][]{Ou2004-vk,Lee2008-mg,Truesdell2006-nj}, several more recent studies found no benefits, even though a plastron was clearly retained on the surface~\citep{Kim2012-iw,Bolognesi2014-vw,Gruncell2014-zp}. \cite{Lee2016-jn} reviewed possible sources of experimental errors that might have affected some of the early measurements. 

\begin{figure}
	\includegraphics[width=\linewidth]{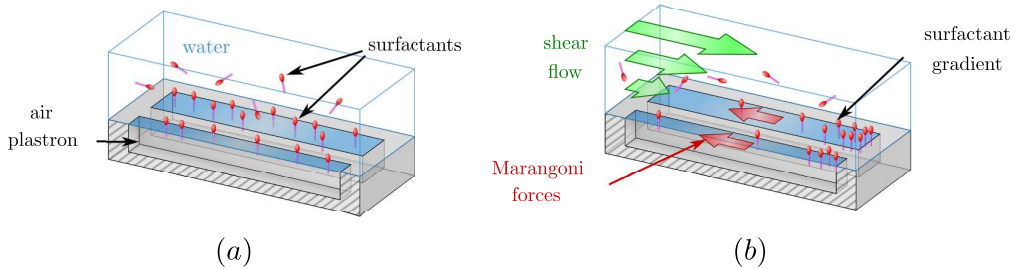}
	\caption{Schematic illustrating the impact of surfactants above a superhydrophobic surface (SHS) made of longitudinal rectangular grating. ($a$) Surfactants present in water adsorb at the air--water interface of the gratings. ($b$) In the presence of an external flow, surfactants distribute in gradients between stagnation points, yielding a Marangoni stress opposing the flow.} 
	\label{fig:surfactantSchematic}
\end{figure}

A key step towards solving this puzzle has come with the realization that surfactants could induce Marangoni stresses that impair drag reduction.
More specifically, \cite{Kim2012-iw} experimentally examined flow over an SHS consisting of gratings perpendicular to the flow, for which they found no measurable slip at the surface. \cite{Bolognesi2014-vw} also found negligible slip for an SHS consisting of gratings aligned with the flow, in contradiction with traditional theoretical {and numerical results}.
\cite{Bolognesi2014-vw} hypothesized that surfactant effects could be to blame.
%
%
Following this hypothesis, surfactants naturally present in water would adsorb onto the air--water interface, as sketched in figure~\ref{fig:surfactantSchematic}($a$). They would then be advected by the flow and  accumulate at downstream stagnation points, where the interface terminates in a three-phase contact line. The resulting surfactant gradient would therefore produce a Marangoni stress opposing the fluid motion, thereby decreasing slip and increasing drag (figure~\ref{fig:surfactantSchematic}$b$). Since traditional models of SHSs are surfactant-free, they cannot account for this additional surfactant-induced Marangoni drag.

Motivated by this hypothesis, \cite{Schaffel2016-mh} performed detailed measurements of the interface slip on an SHS comprising posts. They reported slip velocities far smaller than predicted by surfactant-free simulations. The slip pattern also exhibited strong anisotropy, consistently with what may be expected from surfactant-induced Marangoni stresses in their geometry. Deliberately adding surfactant resulted in a further small decrease in slip, although the magnitude of this change was within experimental uncertainty.
\cite{peaudecerf17} performed unsteady microchannel experiments over SHS consisting of gratings aligned with the flow. By introducing unsteady forcing, they uncovered complex interfacial responses that could only be explained by surfactant effects. They found that, if the driving pressure difference across the microchannel is suddenly removed, the plastron starts flowing backwards relative to the initial flow due to a surfactant-induced Marangoni force. The reverse flow decays as the inverse power of time, consistently with a similarity solution that assumes advection-dominated surfactant transport at the interface.

Since numerous works \citep{Kim2012-iw,Bolognesi2014-vw,Schaffel2016-mh,peaudecerf17,Song2018-uw} observed drastically reduced slip even in nominally clean conditions, \cite{peaudecerf17} performed steady simulations inclusive of surfactants, where they could precisely control surfactant concentrations. They found that surfactant effects can impair drag reduction even at extremely low surfactant concentrations, well below values naturally occurring in the laboratory or the environment. They also found that increasing the streamwise distance between stagnation points on the SHS helped to reduce the surfactant gradient and to increase slip. This explained the large slip achieved in the previous experiments of \cite{Lee2008-mg}, who used a circular rheometer with annular gratings. Annular gratings are effectively infinitely long, without any stagnation point for  surfactants to accumulate, thus avoiding  Marangoni stresses. To illustrate this sensitivity of surfactant-induced Marangoni stresses with respect to the interface geometry, \cite{Temprano-Coleto2018-zu} devised an experiment whereby a complex maze is solved by a small amount of surfactant, which is introduced at the maze entrance. 

More recently, \cite{Song2018-uw} performed detailed experiments on an SHS consisting of a rectangular cavity with small streamwise length. They found that the rectangular gas--liquid interface exhibits recirculation, with reverse flow developing either along the middle or the sides of the plastron, depending on whether the gas--liquid interface is deformed towards the liquid phase (convex) or towards the gas phase (concave), respectively. They performed simulations where a uniform stress was applied to the interface (to approximate a Marangoni stress), showing that the experimentally-observed recirculation pattern could be induced by surfactants. 

While the importance of surfactants is an emerging topic in the context of superhydrophobic surfaces, it should be noted that the importance of surfactant effects has been well-established in many other interfacial flows, often after protracted scientific debates that sought explanations for surprising phenomena. Well-known examples can be found for small bubbles rising in water, where the increased drag due to surfactant adsorption has been studied extensively~\citep[see \eg][and references therein]{bond28,frumkin47,levich62,Palaparthi2006-di}, as well as in dip-coating problems, where accounting for Marangoni stresses is important to predict the coating thickness~\citep{Mayer2012-mc}. In the ocean, the impact of naturally-occurring surfactants is well-established, as they have important effects on wave breaking and gas fluxes \citep{Pereira2018-nw}. Furthermore, steady motions in the bulk (such as internal waves or Langmuir circulations) can cause accumulation of surfactants at the surface. The resulting change in the amplitude of capillary waves affects light scattering, as revealed by satellite photographs~\citep{Kropfli1999-ob}. In laboratory models of oceanic flows, surfactant accumulation can be disproportionately important, driving stresses that qualitatively change the interior flow~\citep{Luzzatto-Fegiz2014-nu}. Traces of surfactant  have also been shown to modify drastically the behavior of the air--water interface of small bubbles probed with atomic force microscopy \citep[see][]{Manor2008,Maali2017}. While a free-slip boundary condition would have been expected, force measurements demonstrated a cross-over between free-slip and no slip depending on the approaching speed of the cantilever or its probing frequency. These modified hydrodynamic boundary conditions are well-modelled by theories that include traces of surfactant, at levels undetectable through traditional surface tension measurements \citep{Manor2008,Maali2017}. These findings further support the notion that surfactant traces can qualitatively alter the hydrodynamics.
%

Predicting surfactant effects is also important since surface-active molecules are inevitably present in both natural end engineered applications. Indeed, biological or environmental samples have been found to contain large amounts of surface-active compounds, including water from seas, rivers, estuaries and fog~\citep{Kropfli1999-ob,Lewis1991-ao,Facchini2000-tu}. For engineered systems,  recently \cite{Hourlier-Fargette2018-tu} used experiments involving insoluble liquid drops in water to demonstrate that uncrosslinked chains of polydimethylsiloxane (PDMS) can act as a surfactant. Since PDMS is one of the most common  materials for microchannel fabrication, their results imply that surfactants are commonly present in microfluidic systems.

While this mounting evidence shows the importance of surfactant effects to superhydrophobic surfaces (at least in the laminar regime), there are presently no theoretical models that can predict slip as a function of surfactant properties and flow geometry.

In this paper, we build a scaling theory that describes slip length, and the associated Marangoni shear stress, in surfactant-laden laminar flows over SHS. As noted earlier, these surfactant effects are induced by accumulation of surfactant at stagnation points on the plastron, which are unavoidable in real applications (except in annular gratings in a rotating flow). As a fundamental model of such a flow, we consider a two-dimensional SHS consisting of transverse grooves, such as those considered by \cite{Kim2012-iw}, \cite{Bolognesi2014-vw} and \cite{Crowdy-2016uk}. This case also serves as an upper bound for the slip and drag reduction that will be obtained in a three-dimensional flow over finite rectangular gratings. Furthermore, the model developed here constitutes a stepping stone towards a more complex theory for three-dimensional flow over SHSs with surfactant.

The problem definition and governing equations are described in \S\ref{sec:probDesc}. 
In \S\ref{sec:slipLength}, we present the key assumptions which  allow us to develop a low-order scaling model for the local slip length at the plastron, as a function of the relevant dimensionless numbers.
In \S\ref{sec:transverseRidges}, a model for the interior flow in a microchannel with a superhydrophobic  side is developed, and coupled to the slip-length model to obtain the effective slip length and drag reduction for the overall channel flow. The overall theory is tested against numerical simulations of the full governing equations. The computational setup is described in \S\ref{sec:numericalsimulations}, and results are reported in \S\ref{sec:results}. Each parameter is varied over several orders of magnitude, confirming each aspect of the theory. The performance, key assumptions and potential uses of the theory are discussed in \S\ref{sec:disc}, with conclusions presented in \S\ref{sec:conc}. To ease adoption and testing of our model, MATLAB codes that automate the theoretical calculations are included as online supplementary material~\cite{githubroutine}.

\begin{figure}
\centering
\includegraphics[width=0.7\linewidth]{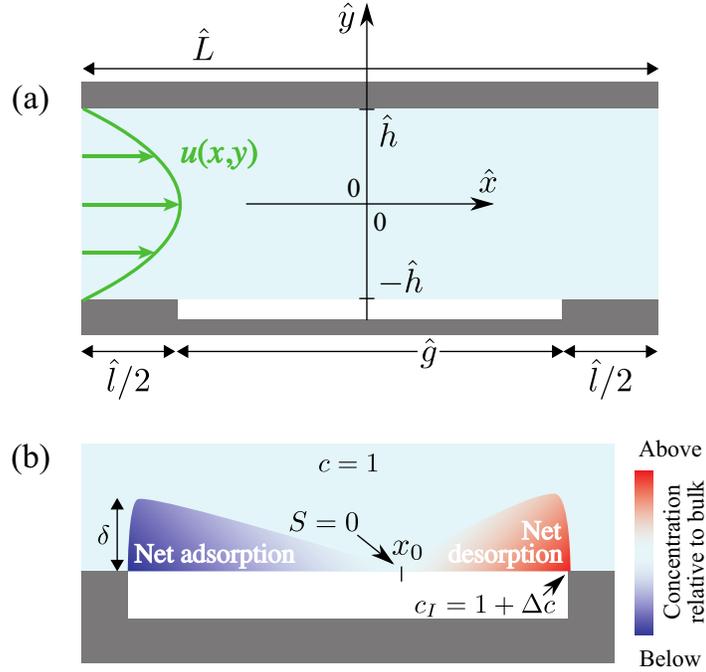}
\caption{ (\textit{a}) Schematic of the geometry of the problem studied. (\textit{b}) Schematic illustrating the bulk concentration profile near the interface. } 
\label{fig:schematicSHSchannel}
\end{figure}

\section{Problem description and governing equations}\label{sec:probDesc}
We study a steady, laminar, two-dimensional liquid flow with a small concentration of surfactant in a channel with a periodic array of flat gas--liquid interfaces on one side, as illustrated in \fig~\ref{fig:schematicSHSchannel}($a$). This geometry is typical of microchannel experiments, where the smooth side of the channel is made transparent to ensure optical access~\citep[see \eg][]{Ou2005-ph,Bolognesi2014-vw,Schaffel2016-mh,peaudecerf17,Song2018-uw}. 
We use hats to denote dimensional quantities throughout the paper, whilst dimensionless quantities are without hats. The dimensional velocity field is $\hat{\boldsymbol{u}}(\hat{x},\hat{y}) = (\hat{u}(\hat{x},\hat{y}),\hat{v}(\hat{x},\hat{y}))$. The surfactant bulk and interfacial concentration fields are $\hat{c}(\hat{x},\hat{y})$ and $\hat{\Gamma}(\hat{x})$, respectively. Owing to the periodicity of the geometry, we can restrict our study to a single periodic cell of total length $\hat{L}$ and height $2\hat{h}$, as shown in figure~\ref{fig:schematicSHSchannel}(\textit{a}). This cell has a centred gas--liquid interface (hereafter designated as ``the interface'') of length $\hat{g}$ at $\hat{y}=-\hat{h}$, with solid surfaces on either side of the interface. The solid surfaces have overall combined length $\hat{l}=\hat{L}-\hat{g}$. Opposite to the interface is a solid surface, located at $\hat{y}=\hat{h}$.
The flow is driven in the positive $\hat{x}$ direction by a constant streamwise mean pressure drop, per unit length, $\hat{G}=-\Delta \hat{p}/\hat{L} >0$. 

We deliberately choose to study the \emph{transverse} flow over SHS gratings of arbitrary but \emph{finite} length $\hat{g}$, instead of the \emph{longitudinal} flow over \emph{infinitely long} gratings as has been done in many previous theoretical and numerical studies \citep{Philip_ZAMP_1972a,Philip_ZAMP_1972b,Lauga_Stone_JFM_2003,sbragaglia07,teo09,Park2013-fj,cheng15,Crowdy-2016uk,Rastegari2018-an}. Indeed, as mentioned in \S1, the establishment of adverse surfactant-induced Marangoni stresses requires stagnation points, as is necessarily found at the end of real SHS gratings, except in the special case of annular gratings \citep{Lee2008-mg}. One of the aim of our model is to predict the effect of $\hat{g}$ on the  effective slip length, following the observations made by \cite{peaudecerf17} that increasing $\hat{g}$ reduces surfactant-induced Marangoni stresses. As also noted earlier, the two-dimensional flow studied here will yield an upper bound for the slip and drag reduction that can be expected in a three-dimensional flow over finite rectangular gratings.


 The governing steady conservation equations for mass, momentum, bulk surfactant, and interfacial surfactant \julien{can be found in dimensional form in \cite{peaudecerf17}. We non-dimensionalize them using the  channel half-height $\hat{h}$ as the length scale, the mean pressure drop per unit length $\hat{G}$ as the scale for pressure gradients, the corresponding velocity $\hat{U} = \hat{G} \hat{h}^2/\hat{\mu}$ as the velocity scale (with $\hat{\mu}$ the dynamic viscosity), the background bulk surfactant concentration $\hat{c}_0$ as the bulk concentration scale, and the maximum packing concentration of the surfactant at the interface $\hat{\Gamma}_m$ \citep{prosser01} as the interfacial concentration scale, such that}
\begin{equation}\label{eq:nondimensionalisation}
    x=\frac{\hat{x}}{\hat{h}},\ y=\frac{\hat{y}}{\hat{h}},\ \bs{u}=\frac{\hat{\bs{u}}}{\hat{U}},\ \nabla p = \frac{\hat{\nabla}\hat{p}}{\hat{G}},\ c=\frac{\hat{c}}{\hat{c}_0},\ \Gamma=\frac{\hat{\Gamma}}{\hat{\Gamma}_m}.
\end{equation}
\julien{The governing equations are, in dimensionless form,}
\begin{eqnarray} 
\nabla \cdot \bs{u} &=& 0, \label{eq:divu}\\  
Re \nabla \cdot (\bs{u}\bs{u})&=& -\nabla{p} + \nabla^2 \bs{u}, \label{eq:u} \\  
\nabla \cdot (\bs{u}c)&=&  \frac{1}{Pe} \nabla^2 c, \label{eq:C} \\    
\fd{}{x}{} \left({u_I}\Gamma\right)&=& \frac{1}{Pe_I} \fd{\Gamma}{x}{2}  + S(c_I,\Gamma)\quad \textrm{on the interface}, \label{eq:Gamma}
\end{eqnarray}
where bold symbols are used for vectors, $u_I(x)$ designates the velocity at the interface, and $p(x,y)$ is the bulk pressure. 
%
The subscript $I$ designates the limit of the bulk quantity considered, as it approaches the interface. In general, this limit is equal to the value taken by the quantity at the interface, except where mentioned explicitly. For instance, we have $u_I(x)  = \lim_{y \to -1^{+}} u(x,y) = u(x,y=-1)$ for $\left|x\right|< g/2$. 
The Reynolds number $Re$, and bulk and interfacial \Peclet numbers $Pe,Pe_I$ are defined below after~(\ref{eq:MarangoniStress}), together with all other dimensionless groups in the problem. A summary is also provided in table~\ref{tab:NDNminmax}.

We assume that the source--sink term modelling the flux of surfactants between the bulk  and the interface follows kinetics consistent with the Frumkin isotherm, which has been found to model accurately single{-}component surfactant systems \citep{Chang_Franses_1995,prosser01},
\begin{equation}\label{eq:S}
S(c_I,\Gamma) = Bi \left( k c_I (1-\Gamma) -  \textrm{e}^{A\Gamma} \Gamma \right).
\end{equation}
with $c_I(x) = \lim_{y \to -1^{+}} c(x,y)$ for $\left|x\right|< g/2$\fernando{. Here} $A$ \fernando{is} the Frumkin interaction parameter\fernando{, which takes negative values $A<0$ for surfactants with attractive intermolecular interactions and positive values $A>0$ in the case of repulsive interactions. This sign convention for $A$ coincides with the one adopted by \cite{prosser01}, but the opposite convention can also be found elsewhere in the literature \citep[\eg in][]{Chang_Franses_1995}}.  In (\ref{eq:S}), we note that the bulk concentration near the interface $c_I$ is different from the interfacial concentration $\Gamma$. This follows the subsurface layer model, where adsorption and desorption kinetics occur between a bulk subsurface layer and the interface \cite[][]{Chang_Franses_1995}. We note that $S>0$ corresponds to an adsorption flux and $S<0$ to a desorption flux, see \fig~\ref{fig:schematicSHSchannel}(\textit{b}). By definition, $\bs{u}$ and $c$ are periodic \fernando{(with period $L$), while the pressure $p$ has a normalized mean drop per unit of length of $-1$, which is enforced by imposing a net pressure drop of value $L$ across each periodic unit of length $L$}\francois{, so that}
\begin{eqnarray}
\boldsymbol{u}(x,y) &=& \boldsymbol{u}(x+L,\,y),\label{eq:period_u}\\
c(x,y) &=& c(x+L,\,y), \quad \label{eq:period_c}\\
\mathrm{and}\;\;p(x,y) &=& p(x+L,\,y) + L.\label{eq:period_p}
\end{eqnarray}
%
The boundary conditions \fernando{include}
\begin{eqnarray} 
\boldsymbol{u} &=& \boldsymbol{0} \quad \textrm{on all solid surfaces \fernando{(no slip)},}\label{eq:BCu}\\
v &=& {0} \quad \textrm{on the interface \fernando{(no penetration)},}\label{eq:BCv}\\
\frac{\partial c}{\partial y} &=& 0 \quad \textrm{on all solid surfaces \fernando{(no flux)},}\\
\fd{ \Gamma}{x}{} &=& 0 \quad \textrm{at}\ (x=\pm \frac{g}{2}\textrm{, }y=-1) \fernando{\textrm{ (no flux)}}. \label{eq:SurfNoFlux}
\end{eqnarray}
\fernando{Additionally, the continuity  of  the  surfactant  fluxes  between  the  bulk  and  the interface is given by}
\begin{equation}
\frac{\chi k}{Pe}\left. \frac{\partial c}{\partial y} \right|_I = S(c_I,\Gamma) \quad \textrm{on the interface.} \label{eq:dCdy} \\    
\end{equation}
\fernando{The last piece of the model is the balance  of  forces  between  the  viscous  drag  from  the  bulk  flow  and the surfactant Marangoni force at the interface. The decrease in surface tension $\sigma$ induced by the surfactant is given by an equation of state consistent with the Frumkin isotherm \citep{prosser01}}\francois{, that is}
\begin{equation}
\sigma = 1 + Ma\,\,Ca \left[ \ln\left(1-\Gamma\right)- \cfrac{A\Gamma^2}{2} \right]\,,\label{eq:eq_state_sigma}
\end{equation}
\fernando{and the Marangoni shear stress at the interface is given by the gradient of surface tension, yielding the last boundary condition} 
\begin{align}
\left.\cfrac{\partial{u}}{\partial{y}}\right|_{I} &= -\cfrac{1}{Ca}\,\,\fd{\sigma}{x}{}\,,\\
\textrm{that  is}\;\;\left.\frac{\partial {u}}{\partial y}\right|_I &=  Ma \left( \frac{1}{1-\Gamma} + A \Gamma \right) \fd{ \Gamma}{x}{} \quad \textrm{on the interface}. \label{eq:MarangoniStress}
\end{align}

\francois{The chosen} characteristic scales imply the following definitions for the dimensionless groups. The Reynolds number is $Re = \hat{\rho}\hat{h}\hat{U}/\hat{\mu}$, with $\hat{\rho}$ the liquid density.
The bulk and interface P\'eclet numbers are $Pe = \hat{h}\hat{U}/\hat{D}$ and $Pe_I = \hat{h}\hat{U}/\hat{D}_I$, where $\hat{D}$ and $\hat{D}_I$ are the bulk and interface surfactant diffusivities, respectively. The Biot number is $Bi = \hat{\kappa}_d \hat{h}/\hat{U}$. The effective bulk concentration is $k = \hat{\kappa}_a \hat{c}_0/\hat{\kappa}_d$, where $\hat{\kappa}_a$ and $\hat{\kappa}_d$ are the adsorption and desorption coefficients, respectively. \paolo{Note that, consistently with the canonical definition of Frumkin kinetics, the adsorption and desorption coefficients $\hat{\kappa}_a$ and $\hat{\kappa}_d$ have different units, so that $k$ is non-dimensional.} The surfactant adsorption--desorption kinetics are parameterized by $\chi = \hat{\kappa}_d \hat{h} / (\hat{\kappa}_a \hat{\Gamma}_m)$. We note that $\chi k=\hat{c}_0\hat{h}/\hat{\Gamma}_m$ in (\ref{eq:dCdy}) is effectively the non-dimensional ratio between the characteristic bulk and interfacial concentration scales. The Marangoni number is $Ma = n_\sigma \hat{R}\hat{T} \hat{\Gamma}_m/(\hat{\mu} \hat{U})$, where $n_\sigma$ is a parameter associated with the Frumkin isotherm \citep{Chang_Franses_1995}, $\hat{R}$ is the universal gas constant, and $\hat{T}$ is the absolute temperature. Temperature-driven Marangoni effects are not considered in this study and we assume that temperature is uniform in the domain. \fernando{Note also that the capillary number $Ca=\hat{\mu}\hat{U}/\hat{\sigma}_0$ (where $\hat{\sigma}_0$ is the surface tension of a clean interface) has no effect in our model, since it does not appear in the final form of the Marangoni boundary condition \eqref{eq:MarangoniStress} and we do not consider any other physical mechanism, such as interface curvature, in which it could play a role (\S\ref{sec:discInterfaceDeformation} provides a discussion of this assumption).}

The governing equations (\ref{eq:divu}--\ref{eq:Gamma}) \fernando{with the periodicity and boundary conditions (\ref{eq:period_u}--\ref{eq:dCdy} and \ref{eq:MarangoniStress})} define a complex nonlinear coupled problem where the unknowns are the two-dimensional velocity field $\bs{u}$, the pressure $p$, the bulk concentration $c$ and the interfacial concentration $\Gamma$.
\paolo{This transport problem depends on nine non-dimensional numbers, which collectively depend on a combination of flow, liquid and surfactant properties, as well as geometry, namely $Re$, $Pe$, $Pe_I$, $Bi$, $k$, $\chi$, $Ma$, $g =\hat{g}/\hat{h}$ and $\gf=g/L = \hat{g}/\hat{L}$. Here $g$ is the normalized interface length, whereas $\gf$ is the gas fraction.} 
According to (\ref{eq:MarangoniStress}), a surfactant-induced Marangoni shear  can develop at the interface when a gradient of interfacial surfactant concentration forms.

The main goal of this study is to determine a low-order model for the interfacial Marangoni shear rate $\left. \partial u/\partial y \right|_I \geq 0$ and the interfacial velocity $u_I\geq 0$ as a function of the nine non-dimensional numbers above, considering realistic parameter regimes. Such model could be used, for instance, to parameterise a slip-length condition in direct numerical simulations of flow over SHS, without having to solve the full complex coupled problem above.

\section{Scaling theory for slip length with surfactant traces}\label{sec:slipLength}
%
%
\subsection{Introducing the Marangoni concentration $k^* \equiv k\,Ma$ for small concentrations}\label{sec:kStar}
The key assumption we propose 
is that the normalised interfacial surfactant concentration $\Gamma$ is sufficiently small{,} such that (\ref{eq:S}) and (\ref{eq:MarangoniStress}) can be linearised. The same assumption was made by \cite{harper04} for the study of air bubbles rising in contaminated water. Hence, we obtain kinetics congruent with the Henry isotherm \citep{Chang_Franses_1995}, namely
\begin{align}
& S(c_I,\Gamma)  \approx Bi \left( k c_I -  \Gamma \right), \label{eq:Slin}\\
& \left. \frac{\partial u}{\partial y}\right|_I \approx  Ma \fd{ \Gamma}{x}{} \quad \textrm{on the interface}. \label{eq:MarangoniStresslin}
\end{align}
In many realistic situations where surfactants are not artificially added, we indeed expect to have low effective bulk concentrations, \ie $k \ll 1$, which generally lead to small interfacial concentrations $\Gamma$. The interfacial concentration is usually away from saturation, \ie $\Gamma\ll 1$, because the maximum packing concentration $\hat{\Gamma}_m$ used in surfactant models is in fact based on geometrical arguments \citep{rosen12} or on achieving good fit of experimental data based on a specific kinetic model \citep{Chang_Franses_1995}. Hence, $\hat{\Gamma}_m$ is usually not attained even when the bulk concentration is at the critical micellar concentration. We also have $A  \lesssim 1$ for common surfactants \citep[see][]{prosser01}. We discuss further the relevance of our assumption $\Gamma \ll 1$ in the context of applications in \S\ref{sec:disc}.

We take advantage of the linearisation of (\ref{eq:S}) and (\ref{eq:MarangoniStress}) to propose a parameter reduction in our problem, by introducing the following rescaled effective Marangoni concentration and surface concentration
\refstepcounter{equation}\label{eq:starqty}
\begin{equation}
k^* \equiv Ma\, k,\qquad \Gamma^* \equiv Ma\, \Gamma. 
 \tag{\theequation \textit{a,b}}\label{eq:starqtya--d}
\end{equation}
Substituting $k=k^*/Ma$ and $\Gamma=\Gamma^*/Ma$ into (\ref{eq:Gamma}), (\ref{eq:dCdy}) and (\ref{eq:SurfNoFlux}), with the Henry kinetics (\ref{eq:Slin}) and (\ref{eq:MarangoniStresslin}), we obtain a set of equations where $k$ and $Ma$ have been combined to form $k^*$, thereby reducing by one the number of dimensionless groups. This can be verified by examining the updated version of the complete set of equations (\ref{eq:divu})-(\ref{eq:SurfNoFlux}), which becomes
\begin{eqnarray} 
\nabla \cdot \bs{u} &=& 0, \label{eq:divuStar}\\  
Re\nabla \cdot (\bs{u}\bs{u})&=& -\nabla{p} + \nabla^2 \bs{u}, \label{eq:uStar} \\  
\nabla \cdot (\bs{u}c)&=&  \frac{1}{Pe} \nabla^2 c, \label{eq:Cstar} \\    
\fd{}{x}{} \left({u_I}\Gamma^*\right)&=& \frac{1}{Pe_I} \fd{\Gamma^*}{x}{2}  + Bi(k^*c_I-\Gamma^*)\quad \textrm{on the interface}, \label{eq:GammaStar}
\end{eqnarray}
with boundary conditions
\begin{eqnarray}
\boldsymbol{u} &=& \boldsymbol{0} \quad \textrm{on all solid surfaces},\label{eq:BCuStar}\\
v &=& {0} \quad \textrm{on the interface},\label{eq:BCvStar}\\
\frac{\partial c}{\partial y} &=& 0 \quad \textrm{on all solid surfaces},\\
\fd{ \Gamma^*}{x}{} &=& 0 \quad \textrm{at}\ (x=\pm \frac{g}{2},y=-1), \label{eq:SurfNoFluxStar}\\
\frac{\chi k^*}{Pe}\left. \frac{\partial c}{\partial y} \right|_I  &=& Bi (k^*c_I-\Gamma^*) \quad \textrm{on the interface,} \label{eq:dCdyStar}  \\
\left.\frac{\partial {u}}{\partial y}\right|_I &=&  \fd{ \Gamma^*}{x}{} \quad \textrm{on the interface}, \label{eq:MarangoniStressStar}
\end{eqnarray}
such that the three quantities $k$, $\Gamma$ and $Ma$ have been replaced by the dimensionless number  $k^*$ and the variable $\Gamma^*$.

\subsection{Scaling theory for surfactant dynamics}\label{sec:scalingTheory}
To make further progress in modelling the  shear rate $\left. \partial u/\partial y \right|_I $ and  velocity $u_I$, we perform a scale analysis on the  equations in our problem, starting with rearranging (\ref{eq:dCdyStar}), which expresses continuity of surfactant fluxes between the bulk and the interface
\begin{equation}\label{eq:dcdy_start}
\left. \frac{\partial c}{\partial y} \right|_I  = \frac{Bi\,Pe}{\chi k^*} \left(k^* c_I - \Gamma^* \right).
\end{equation}
For steady flows, adsorption and desorption fluxes between the bulk and the interface are in balance overall, implying 
\begin{equation}
    \int_{-g/2}^{g/2} \left. \frac{\partial c}{\partial y} \right|_I \mathrm{d}x=  \frac{Bi\,Pe}{\chi k^*} \int_{-g/2}^{g/2} \left(k^* c_I - \Gamma^* \right) \mathrm{d}x=0,
    \label{eq:ads_des_SS}
\end{equation}
such that, by the mean value theorem, there is a point on the interface where $\left. {\partial c}/{\partial y} \right|_I = 0$ and $k^* c_I = \Gamma^*$.
With a flow in the positive $x$-direction, interfacial surfactant $\Gamma^*$ is advected downstream, such that the beginning of the interface has a lower surfactant concentration, implying that $\Gamma^*<k^* c_I$, and that an adsorption flux exists from the bulk onto the beginning of the interface, such that $\left. \partial c/\partial y \right|_I > 0$ there, as illustrated in figure~\ref{fig:schematicSHSchannel}(\textit{b}). 
By the same argument, near the end of the interface, a higher surfactant concentration $\Gamma^*> k^* c_I$ leads to desorption from the interface into the bulk, implying $\left. \partial c/\partial y \right|_I < 0$. Therefore, somewhere along the interface, we must have $\left. \partial c/\partial y \right|_I = 0$. We designate by $x_0$ this location where the kinetics flux $S=0$, as depicted in figure~\ref{fig:schematicSHSchannel}(\textit{b}).

In addition, at the beginning of the interface, $c_I$ is less than the bulk concentration, \ie $c_I<1$ with our nondimensionalization, whereas towards the end of the interface, where surfactants accumulate, $c_I > 1$. This means that, at a specific location along the interface, the concentration near the interface is equal to the background bulk concentration, that is $c_I=1$. 
Taking $c_I\sim 1$ along the interface, we then find that (\ref{eq:dcdy_start}) implies that the interfacial concentration scales as $\Gamma^*\sim k^*$. 

Next, assuming that the  variations of $c_I$ and $\Gamma$ scale in the same way  for the adsorption region, $-g/2<x<x_0$, and the desorption region, $x_0<x<g/2$
, we have
\begin{eqnarray}\label{eq:concexpansion}
{\Gamma^*} \sim k^* \mp \Delta \Gamma^*, \quad c_I \sim 1 \mp \Delta c_I,
\end{eqnarray}
for the adsorption ($-$) and desorption ($+$) regions, respectively (see \fig~\ref{fig:schematicSHSchannel}\textit{b}). The quantities $\Delta \Gamma^*$ and $\Delta c_I$ are the characteristic variations of $\Gamma^*$ and $c_I$, respectively. We must have $\Delta \Gamma^*> k^*\Delta c_I > 0$ to satisfy the direction of the kinetics flux, as described above.

From the relation between Marangoni stress and surfactant gradient (\ref{eq:MarangoniStresslin}), we also have
\begin{equation}\label{eq:ScalingGammaMa}
\Delta \Gamma^* \sim g\gamma_{Ma},
\end{equation}
where 
\begin{equation}\label{eq:gammaMadef}
\gamma_{Ma}=\frac{1}{g}\int_{-g/2}^{g/2} \left. \fp{u}{y}{}\right|_I \mathrm{d}x
\end{equation} 
is the average shear rate induced by Marangoni stresses along the interface, such that $\gamma_{Ma}=0$ corresponds to free-slip at the interface and $\gamma_{Ma}=1$ corresponds to a no-slip interface. Then, a scale analysis of (\ref{eq:dcdy_start}) gives
\begin{equation}\label{eq:SurfactantFluxScalingKineticsBulk}
\left.\frac{\partial c}{\partial y} \right|_I\sim \frac{\Delta c_I}{\delta}\sim \frac{Bi\,Pe}{\chi k^*}\left( g \gamma_{Ma}- k^*\Delta c_I \right),
\end{equation}
where $\delta$ is the typical thickness of the diffusive  layer of bulk surfactant.
To estimate $\delta$, we can use the bulk advection--diffusion equation (\ref{eq:C}). At high P\'eclet numbers, $Pe\gg 1$, the diffusive layer of surfactant forms a thin  boundary layer. As explained in detail in appendix~\ref{apdx:DiffBLthickness}, there are two main asymptotic regimes depending on whether there is slip or not at the interface. For large slip and small interfacial shear rate, $\gamma_{Ma}\ll 1$, we can show that the boundary layer thickness scales as (see appendix~\ref{apdx:DiffBLthickness})
\begin{eqnarray}
    \frac{\delta}{g} & = & \delta_{0,1} \left(1 + \delta_{1,1} g^2 Pe\right)^{-1/2}\quad \textrm{for}\ g \lesssim 1, \label{eq:deltaPe12}\\
    \frac{\delta}{g} & = & \delta_{0,2} \left(1 + \delta_{1,2} g Pe\right)^{-1/2} \quad \textrm{for}\ g \gtrsim 1, \label{eq:deltagPe12}
\end{eqnarray}
where $\delta_{0,1}$, $\delta_{1,1}$, $\delta_{0,2}$, $\delta_{1,2}$ are empirical parameters which need to be determined. We note that the scaling $\delta\sim Pe^{-1/2}$ at large \Peclet numbers corresponds to having a uniform velocity in the diffusive boundary layer, consistently with the case $\gamma_{Ma}\ll 1$. 

For negligible slip at the interface and $\gamma_{Ma}\sim 1$, we obtain
\begin{equation}\label{eq:deltagPe13}
    \frac{\delta}{g} = \delta_{0,3} \left(1 + \delta_{1,3} g^2 Pe\right)^{-1/3},
\end{equation}
for any $g>0$, and with $\delta_{0,3}$ and $\delta_{1,3}$ two empirical parameters which need to be determined. This corresponds to the \Leveque regime \citep{leveque28,landel16}, giving a power law $\delta\sim Pe^{-1/3}$ at large \Peclet numbers owing to a linear shear rate profile in the diffusive boundary layer.
The scalings (\ref{eq:deltaPe12})--(\ref{eq:deltagPe13}) assume that: (i) the variation of the bulk concentration along the interface is sufficiently smooth; (ii) the boundary layer is not confined vertically, \ie $\delta\lesssim 1$; and (iii) the diffusive boundary layers between consecutive interfaces are independent. As we will discuss in \S\ref{sec:results}, our scaling prediction remains accurate even for confined diffusive boundary layers $\delta\sim 1$.

With $\delta$ assumed known in terms of $g$ and $Pe$, we rearrange (\ref{eq:SurfactantFluxScalingKineticsBulk}) to solve for $\Delta c_I$
\begin{equation}
\Delta c_I \sim   \gamma_{Ma} \frac{\frac{Bi\,Pe}{\chi k^*}\,g\, \delta }{1 + \frac{Bi\,Pe}{\chi}\delta},
\end{equation}
such that, dividing by $\delta$, we obtain a scaling that relates the kinetics flux to the shear
\begin{equation}\label{eq:SurfactantFluxScalingKineticsBulk2}
\left.\frac{\partial c}{\partial y} \right|_I\sim \frac{\Delta c_I}{\delta} \sim \gamma_{Ma} \frac{\frac{Bi\,Pe}{\chi k^*} g }{1 + \frac{Bi\,Pe}{\chi}\delta}.
\end{equation}

\subsection{Scaling for the interfacial velocity and for the slip length}

We now seek a scaling expression for  $u_I$. We integrate the interfacial advection--diffusion equation (\ref{eq:GammaStar}) from the upstream stagnation point $x=-g/2$ to $x_0$. We find 
\begin{equation}\label{eq:uIGamma}
   \left.  \left(u_I\Gamma^*\right)\right|_{x_0} =  \left.\frac{1}{Pe_I}\fd{\Gamma^*}{x}{}\right|_{x_0} +  \frac{k^*\chi}{ Pe} \int_{-g/2}^{x_0} \left.  \fp{c}{y}{}\right|_{I} \mathrm{d}x,
\end{equation}
where we used the no-slip boundary condition (\ref{eq:BCuStar}) at $x=-g/2$ for the left hand side,  the no flux boundary condition (\ref{eq:SurfNoFluxStar}) at $x=-g/2$ for the first term on the right hand side, as well as the continuity of flux condition (\ref{eq:dCdyStar}) for the last term. To write the right-hand side in terms of $\gamma_{Ma}$, note that $\left. \Gamma^*\right|_{x_0}\sim k^*$ and $\left.\mathrm{d}\Gamma^*/\mathrm{d}x\right|_{x_0}\sim \gamma_{Ma}$. For the last term, we use (\ref{eq:SurfactantFluxScalingKineticsBulk2}) to scale the integral
\begin{equation}
 \int_{-g/2}^{x_0} \left.  \fp{c}{y}{}\right|_{I} \mathrm{d}x \sim g \frac{\Delta c_I}{\delta}  \sim \gamma_{Ma} \frac{\frac{Bi\,Pe}{\chi k^*} g^2 }{1 + \frac{Bi\,Pe}{\chi}\delta}.
\end{equation}
Substituting into (\ref{eq:uIGamma}) we obtain a scaling relation between interfacial velocity and shear. Introducing empirical prefactors (to be determined) ahead of each term, we write
%
%
\begin{equation}\label{eq:scalinguIsurfactant}
\left. u_I\right|_{x_0} =  \frac{2}{a_1}\frac{1}{k^*}\left(\frac{1}{Pe_I} + a_2 \frac{g^2 Bi}{1+\frac{BiPe}{\chi}\delta} \right) \gamma_{Ma},
\end{equation}
where $a_1, a_2$ are empirical parameters; the choice of writing $2/a_1$ for the overall prefactor leads to a more convenient expression for the results later in \S\ref{sec:EffectiveSlipLength}. 

As also noted in the previous section, scaling expressions for the boundary layer thickness $\delta$ are given by (\ref{eq:deltaPe12}), (\ref{eq:deltagPe12}) or (\ref{eq:deltagPe13}), which depend on $g$, $\gf$ and $\gamma_{Ma}$. Therefore, our scaling is a nonlinear function of the Marangoni shear rate. However, in the comparison of our model with numerical simulations (see \S\ref{sec:results}), we find that the nonlinear dependence of $\gamma_{Ma}$ with $\delta$ is actually weak. Consequently, we adopt only (\ref{eq:deltagPe13}) in our model. This proves to be a good approximation and allows us to regard $\delta$ as independent from $\gamma_{Ma}$.

Furthermore, we note that a first-order linear expansion of the concentrations $c$ and $\Gamma$ near $x_0$ predicts $a_2=1/8$ since $x_0=0$ (that is, $x_0$ is at the mid-gap location) due to the balance of desorption and adsorption fluxes along the interface. 

A characteristic scale for the slip length near $x_0$, which corresponds to the mid-gap of the interface under our assumptions, is therefore simply
\begin{equation}\label{eq:sliplengthmidgap}
\lambda_{x_0} = \frac{\left. u_I\right|_{x_0}}{\gamma_{Ma}} = \frac{2}{a_1}\frac{1}{k^*}\left(\frac{1}{Pe_I} + a_2 \frac{g^2 Bi}{1+\frac{BiPe}{\chi}\delta} \right).
\end{equation}
This scaling prediction shows that the local slip length $\lambda_{x_0}$ depends strongly on the Marangoni concentration $k^*=k\,Ma$ and the normalised gap length $g$. It is intuitive that increasing the gap length tends to increase the slip length, since it would reduce the concentration gradient at the interface and thus the  opposing Marangoni stress. In contrast, increasing the effective bulk surfactant concentration $k$ or the Marangoni number tends to reduce the slip length, as expected. We also find increasing the bulk or interfacial \Peclet numbers, $Pe$ or $Pe_I$, reduces $\lambda_{x_0}$. Increasing the Biot  or $\chi$ numbers has a positive effect on the slip length.

However, we note that (\ref{eq:sliplengthmidgap}) is only a local measure of the characteristic slip length near the middle of the interface, where $S(x_0)=0$. In order to have an effective or global slip length over the entire SHS which takes into account all interfaces and  solid ridges, we also need to model the channel flow over the SHS. In the next section, we analyse the remaining governing equations for the flow, \ie the continuity and Navier--Stokes equations (\ref{eq:divuStar}) and (\ref{eq:uStar}), to study how the flow is affected by a SHS with a surfactant-induced Marangoni stress over the interfaces.

\section{Complete model for effective slip in channel flows with one-sided periodic transverse ridges}\label{sec:transverseRidges}


\subsection{Stokes flow model for SHS channels with surfactant contamination}\label{sec:StokesFlowModel}
According to equation (\ref{eq:MarangoniStresslin}), interfacial surfactant concentration gradients can generate a Marangoni shear rate at the interface $\left. \partial u/\partial y \right|_I \approx \textrm{d}\Gamma^*/\textrm{d}x \geq 0$. In this section, we derive an expression for how interfacial stresses with arbitrary profile can affect the flow over a periodic SHS. The geometry follows the same schematic presented in figure~\ref{fig:schematicSHSchannel}.
Such a periodic SHS  arrangement was studied in detail by \cite{Lauga_Stone_JFM_2003} and \cite{teo09} for  a shear-free interface, \ie with $\left. \partial u/\partial y \right|_I=0$ along the interface, at low Reynolds number. Here we generalize their approach to also study the case where $\left. \partial u/\partial y \right|_I\geq 0$. We also assume $Re\ll 1$, such that (\ref{eq:uStar}) simplifies to the Stokes flow equation
\begin{equation}\label{eq:stokesflow}
\nabla{p} = \nabla^2 \bs{u}.
\end{equation}
Taking the curl of (\ref{eq:stokesflow}) and using the continuity equation (\ref{eq:divuStar}), we find that the pressure field $p$ and the vorticity field, $\bs{\omega}=\nabla \times \bs{u}$ are both solutions of Laplace's equation.
Using the superposition principle to solve Laplace's equation for the vorticity, we decompose it as the sum of the two-dimensional Poiseuille flow component, which is a pressure driven flow in a channel with full solid walls on both sides (denoted by a subscript $p$), and a deviating component (denoted by a subscript $d$), such that
\begin{equation}
\omega = \omega_p + \omega_d,
\end{equation}
where $\omega_p=y$. As the flow is incompressible, we can also use the streamfunction $\bs{\Psi}$, defined such that $\bs{u}= \nabla \times \bs{\Psi}$, and which is the solution of the biharmonic equation $\nabla^4 \bs{\Psi} =\bs{0}$. Note that $\bs{\Psi} = (0,0,\Psi)$ for two-dimensional flows. The solution for the deviating component of the vorticity is obtained using separation of variables considering the periodicity of the flow with wavelength $L$. Integrating twice, we then obtain the deviation streamfunction \citep{Lauga_Stone_JFM_2003,teo09}. Noting that the mean pressure gradient imposed by the deviating field is zero, neglecting the constant of integration, and using the  no-flow boundary condition $v=0$ in (\ref{eq:BCuStar}) and  (\ref{eq:BCvStar}), the deviating component of the streamfunction is
\begin{align} \label{eq:genPsid}
\Psi_d = & B\frac{y^2}{2} + E y + \sum_{n=1}^{\infty}\Big\{e_n \big[\cosh(k_n y)-\coth(k_n)y \sinh(k_n y) \big]  \nonumber\\
&   + d_n\big[\sinh(k_n y) - \tanh(k_n) y \cosh(k_n y) \big]  \Big\} \cos(k_n x),
\end{align}
where $k_n=2\pi n/L$, and $B$, $E$, $e_n$ and $d_n$ are unknowns to be determined using the other boundary conditions. The streamfunction for the Poiseuille component is 
\begin{equation} \label{eq:Psip}
\Psi_p = \frac{1}{6}(3y-y^3).
\end{equation}
Up to this point,  (\ref{eq:genPsid}) and (\ref{eq:Psip}) are general solutions for any arrangement and geometry of SHS in a two-dimensional Stokes flow channel: \ie they are not limited to one-sided SHS,  symmetric patterns, or  a particular  shear rate profile at the interface.

With our geometry, using the no-slip boundary condition on the solid wall side at $y=1$ for all $x$, we find $B=-E$ and
\begin{equation}
g_n = \frac{e_n}{d_n} =  -\frac{\sinh(k_n)-k_n\cosh(k_n)+k_n \tanh(k_n)\sinh(k_n)}{\cosh(k_n)-k_n\sinh(k_n)+k_n\coth(k_n)\cosh(k_n)}.
\end{equation}
Hence, the deviating streamfunction simplifies to
\begin{align} \label{eq:Psidsimp}
\Psi_d = & \left(-\frac{y^2}{2} +  y\right)E + \sum_{n=1}^{\infty}d_n\Big\{ g_n\bigl[ \cosh(k_n y)-\coth(k_n) y \sinh(k_n y) \bigr] \nonumber\\
&  +  \sinh(k_n y) - \tanh(k_n) y \cosh(k_n y)  \Big\} \cos(k_n x).
\end{align}

To determine the unknowns $E$ and $d_n$ for $n\geq 1$, we can use the  no-slip boundary condition on the SHS side. At  $y=-1$ for $g/2< \left|x\right| < L/2$, we have the condition
\begin{equation}\label{eq:monster1}
0 = 2E + \sum_{n=1}^{\infty} d_n \alpha_n \cos(k_n x),
\end{equation}
where
\begin{equation} \label{eq:alpha_n}
\alpha_n = 2 k_n \big[\cosh(k_n)-\tanh(k_n)\sinh(k_n) \big] - 2\sinh(k_n).
\end{equation}
We then apply the last boundary condition  on the interface, where we assume that there is an arbitrary  shear rate profile $\left.\partial u/\partial y\right|_I (x) \geq 0$. 
Hence,  we obtain the general condition 
\begin{equation}\label{eq:BCinterface}
0 = \left.\frac{\partial u}{\partial y}\right|_I -1 + E + \sum_{n=1}^{\infty} d_n \beta_n \cos(k_n x)
\end{equation}
for $\left|x\right| < g/2$, $y=-1$, and with
\begin{equation} \label{eq:beta_n}
\beta_n = 2k_n \big[ g_n \coth(k_n)\cosh(k_n) - \tanh(k_n)\sinh(k_n)\big].
\end{equation}

To make further progress and obtain a relationship between the interfacial shear rate and the interfacial velocity, we now assume that the interfacial shear rate is uniform along the interface: $\left.\partial u/\partial y\right|_I=\gamma_{Ma}$, where $0\leq \gamma_{Ma}\leq 1$ corresponds to the interface-averaged surfactant-induced Marangoni shear rate, as introduced previously in (\ref{eq:ScalingGammaMa}). \julien{This assumption is consistent with having a uniform concentration gradient, following the linearised coupling condition (\ref{eq:MarangoniStresslin}). In the context of air bubbles rising in surfactant-contaminated water, this assumption is also consistent with the `uniformly retarded' regime described for instance by \cite{Palaparthi2006-di}. In the context of SHS, a similar assumption was made by \cite{schonecker13}  to model viscous effects from a gas phase trapped inside the cavities of the SHS. This allowed them to decouple the flow above the interface from the flow in the cavity of the SHS.} We discuss further the relevance of this assumption in applications in \S\ref{sec:disc}.
Hence, (\ref{eq:BCinterface}) becomes
\begin{equation}\label{eq:monster2}
0 = \gamma_{Ma} -1 + E + \sum_{n=1}^{\infty} d_n \beta_n \cos(k_n x).
\end{equation}
If ${\gamma}_{Ma}=0$ in the equation above, the interface is stress free and the surfactant concentration gradient  at the interface vanishes. The surface is completely immobilized if ${\gamma}_{Ma}=1$, and the flow follows a channel Poiseuille flow.

Following \cite{Lauga_Stone_JFM_2003}, we can compute an approximation of the solution by truncating the series in equations (\ref{eq:monster1}) and (\ref{eq:monster2}) at \(n=N-1\), multiplying (\ref{eq:monster1}) and (\ref{eq:monster2}) by $\cos(2 \pi m r)$ for \(m \in [0,N-1]\) (with $r=x/L$) and integrating them for \(r \in (\gf/2,1/2)\) and \(r \in (0,\gf/2)\), respectively, where $\gf$ is the gas fraction. Summing together the results for each \(m\) in one single equation, we finally obtain a linear system of \(N\) equations for the \(N\) unknown coefficients \(E\) and \(d_n\) for \(n \in [1,N-1]\), which we can solve numerically.
The linear system in matrix form is, for  $m \in [0,N-1]$ and $n \in [0,N-1]$,
\begin{equation}\label{eq:linsysmatrix}
\bs{A}_{m,n} \bs{U}_n = \bs{B}_m,
\end{equation}
with $U_0 = E$ and $U_n$ = $d_n$. The square matrix $\bs{A}_{m,n} $ has coefficients
\begin{align}
A_{0,0} = & 1-\frac{\gf}{2}, \\
A_{0,n} = & (\beta_n-\alpha_n) \frac{\sin(\pi n \gf)}{2\pi n}, \ n>0 \\
A_{m,0} = & -\frac{\sin(\pi m \gf)}{2\pi m}, \ m>0 \\
A_{n,n} = & \frac{\alpha_n}{4} + (\beta_n-\alpha_n) \left(\frac{\gf}{4} + \frac{\sin(2 \pi n \gf)}{8\pi n} \right), \ n>0 \\
A_{m,n} = & (\beta_n-\alpha_n) \frac{1}{4\pi} \left( \frac{\sin( \pi (m+n) \gf)}{m+n} + \frac{\sin( \pi (m-n) \gf)}{m-n} \right), \ m\neq n>0 
\end{align}
and the vector $\bs{B}_{m} $ has coefficients
\begin{align}
B_{0} = & (1-{\gamma}_{Ma}) \frac{\gf}{2}, \label{eq:B0}\\
B_{m} = & (1-{\gamma}_{Ma}) \frac{\sin(\pi m \gf)}{2\pi m},  \ m>0. \label{eq:Bm}
\end{align}
Care must be taken at large $n$, where the system is not well conditioned, as pointed out by \cite{teo09}. We provide, as supplementary material, MATLAB routines solving the linear system (\ref{eq:linsysmatrix})~\cite{githubroutine}.

\subsection{Interfacial slip velocity}

Once all the coefficients $E$ and \(d_n\) are computed, the non-dimensional slip velocity at the interface $u_I$ can be determined to machine precision, depending on the size $N$ of the matrix $\bs{A}$, such that
\begin{equation}\label{eq:uIallx}
	u_I = 2E +  \sum_{n=1}^{\infty} d_n \alpha_n \cos(k_n x).
\end{equation}
Through its coefficients $E$ and $d_n$, (\ref{eq:uIallx}) is a function of the uniform Marangoni interfacial shear rate ${\gamma}_{Ma}$ and of the two non-dimensional geometrical parameters $g$ and $\gf$. Hence, we have
\begin{equation}
u_I = \mathcal{G}({\gamma}_{Ma},g,\gf,x).
\end{equation}
The function $\mathcal{G}$ is only known implicitly through the solution of the linear system (\ref{eq:linsysmatrix}). In practice, it would be useful to obtain an explicit analytical solution, or at least a scaling expression for $\mathcal{G}$ which can give an approximate solution to the  coupled surfactant--flow transport problem in combination with (\ref{eq:scalinguIsurfactant}). 
In the linear system (\ref{eq:linsysmatrix}), we can factorize all the coefficients of $\bs{B}_{m} $ by $(1-{\gamma}_{Ma})$. This means that $E$ and $d_n$  are proportional to  $(1-\gamma_{Ma})$ for all $n\geq1$. Thus, the velocity at the interface is such that
\begin{equation}\label{eq:uIscaled1}
u_I = 2(1-{\gamma}_{Ma}) \mathcal{F}(g,\gf,x),
\end{equation}
where, again, $\mathcal{F}$ is an implicit function. Now, $\mathcal{F}$ is decoupled from the surfactant transport problem since it does not depend on $\gamma_{Ma}$. It can thus be computed to arbitrary numerical precision for each couple of geometrical non-dimensional parameters $(g,\gf)$ and for all $x$ by solving the linear system (\ref{eq:linsysmatrix}) in the surfactant-free case, \ie setting $\gamma_{Ma}=0$ in (\ref{eq:B0}) and (\ref{eq:Bm}). 

\begin{figure}
\centering
\includegraphics[width=0.95\linewidth]{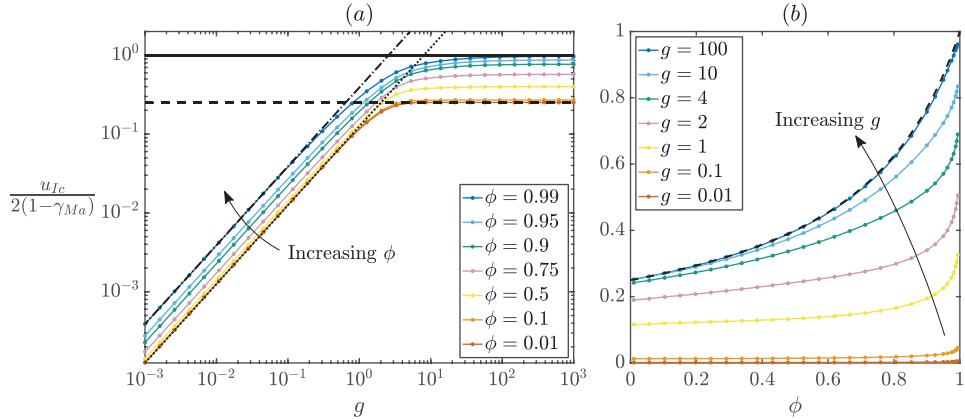}
\caption{Variation of the normalised mid-gap interfacial velocity, $u_{Ic}/(2(1-\gamma_{Ma}))=\mathcal{F}(g,\gf,x=0)=\mathcal{F}_0(g,\gf)$ (see (\ref{eq:uIscaled1}) and text), as a function of: $(a)$ the non-dimensional interfacial length $g$, for different non-dimensional gas fraction from $\gf=0.01$ to $0.99$ (shown with different colors, see legend); $(b)$ the gas fraction $\gf$, for various interfacial lengths $g$ (shown with different colors, see legend). The implicit function $\mathcal{F}_0(g,\gf)$ has been computed by solving the linear system (\ref{eq:linsysmatrix}) with $N=500$, except in the more demanding cases of $0.01<\gf\leq{0.1}$ ($N=2{,}500$), $0.99\leq\gf\leq{1}$ ($N=2{,}500$) and $0\leq\phi\leq{0.01}$ ($N=15{,}000$). In (\textit{a}) the black dotted line is plotted using (\ref{eq:uicScaling1}) for $\gf\ll 1$ and $g\lesssim 1$, the black dot-dashed line is plotted using \eqref{eq:uIcenter_for_k_inf_Sbrag} for  $g\lesssim 1$ and $\gf= 0.99$, and the black dashed line is plotted using (\ref{eq:uicScaling2}) for $\phi\ll 1$ and $g\gtrsim 1$. The black solid line corresponds to the maximum asymptotic value for $\gf\to 1$: $u_{I,c}\to u_u(y=-1)=2(1-\gamma_{Ma})$ (see (\ref{eq:uuniformasym})). In (\textit{b}), the black dashed line has been plotted using the asymptotic trend (\ref{eq:uicScaling3}) for $g\gg 1$ (see also \App~\ref{apdx:Asympt_Vel_Plastron}). } 
\label{fig:uIc_with_g_and_phi}
\end{figure}

Based on this observation, $\mathcal{F}(g,\gf,x) = u_{I}(x)/(2(1-\gamma_{Ma}))$ is a normalised interfacial velocity.
In figure~\ref{fig:uIc_with_g_and_phi}(\textit{a}), we plot on a log--log scale this normalised interfacial velocity at the middle of the gap, $x=0$: 
\begin{equation}\label{eq:uicScaling0}
\frac{u_{Ic}}{2(1-\gamma_{Ma})}=\mathcal{F}(g,\gf,x=0)=\mathcal{F}_0(g,\gf),
\end{equation} 
as a function of $g$ and for different $\gf$ (shown with different colors, see legend).  For $\gf\ll 1$ and $g\lesssim 1$, the normalised interfacial velocity follows a linear asymptotic trend 
\begin{equation}\label{eq:uicScaling1}
\frac{u_{Ic}}{2(1-{\gamma}_{Ma})} \simeq \frac{g}{8},
\end{equation} 
plotted with a black dotted line in figure~\ref{fig:uIc_with_g_and_phi}(\textit{a}). We can see that for  $\gf=0.99$ the interfacial velocity still follows a linear scaling, although with a higher slope than in the asymptotic limit (\ref{eq:uicScaling1}), as shown by the black dot-dashed line in \fig~\ref{fig:uIc_with_g_and_phi}(\textit{a}), which was computed using \eqref{eq:uIcenter_for_k_inf_Sbrag}.
At large gap length, $g\gg 1$, and for low gas fraction, $\gf\ll 1$, the interfacial velocity collapses on the asymptotic plateau 
\begin{equation}\label{eq:uicScaling2}
\frac{u_{Ic}}{2(1-{\gamma}_{Ma})} \to \frac{1}{4},
\end{equation} 
plotted with a black dashed line in figure~\ref{fig:uIc_with_g_and_phi}(\textit{a}). More details about the behaviour of the interfacial velocity $u_{Ic}$ with $\gf$ and $g$ and  the two asymptotic limits (\ref{eq:uicScaling1}) and (\ref{eq:uicScaling2}) can be found in \App~\ref{apdx:Asympt_Vel_Plastron}. The transition observed at $g\sim 1$ from a linear trend towards a plateau is due to the importance of the opposite wall at $y=1$ through viscous effects. 

We note that the behaviour of $u_{I,c}/(2(1-\gamma_{Ma}))$ is similar across all $g$ and for any $\gf$.  This function goes from a linear behaviour for $g\lesssim 1 $ to a plateau for $g\gtrsim 1 $, and with simple asymptotics in the case $\gf\ll 1$. Most of the data in figure~\ref{fig:uIc_with_g_and_phi}(\textit{a}) follows these limiting regimes, suggesting that asymptotic results are sufficiently accurate in many applications.

This common behavior of the interfacial velocity might also suggest that the velocity field follows a closed analytical form. However, we have not been able to demonstrate this theoretically from the biharmonic equation. As far as we are aware, the case of Stokes flow in a transverse channel with mixed boundary conditions changing twice (on one or both channel sides), which is reminiscent of the longitudinal-channel work of \cite{Philip_ZAMP_1972a}, has not been shown to have a closed analytical form in the literature. It would be valuable to re-examine the present problem with conformal mapping tools similar to those used by~\cite{Crowdy-2016uk,Crowdy2017-pc}.

Figure~\ref{fig:uIc_with_g_and_phi}(\textit{b}) plots curves of $u_{I,c}/(2(1-\gamma_{Ma}))$ versus gas fraction $\gf$, with $g$ as a parameter. As the gas fraction $\gf$ increases towards $1$, the normalised interfacial velocity increases rapidly at any fixed $g$.
In the limit $\gf\to 1$ we have  
\begin{equation}\label{eq:uIcasym1}
\frac{u_{Ic}}{2(1-{\gamma}_{Ma})} \to 1,     
\end{equation}
which can be predicted from the velocity field with uniform boundary conditions at the top and bottom sides, that is $u(y=1)=0$ and $\textrm{d}u_u/\textrm{d}y(y=-1)={\gamma_{Ma}}$, respectively. The solution to the Stokes problem (\ref{eq:stokesflow}) with these uniform boundary conditions is independent of $x$:
\begin{equation}\label{eq:uuniformasym}
u_u = \frac{1}{2}\left(1- y^2\right) + (1-{\gamma_{Ma}}) (1-y),
\end{equation} 
and can be used to yield the limit of $u_{Ic}\to u_{u}(y=-1)$ for $\gf\to 1$. At $g\lesssim 1$, it is also clear from \fig~\ref{fig:uIc_with_g_and_phi}(\textit{b}) that $u_{Ic}/(2(1-\gamma_{Ma}))\to 1$ only for gas fraction very close to 1, \ie in the limit $\gf\to 1$, as already observed in \fig~\ref{fig:uIc_with_g_and_phi}(\textit{a}). This result confirms the range of validity of the first scaling (\ref{eq:deltaPe12}) for the diffusive boundary layer thickness $\delta$.
Then, we can show (see \App~\ref{apdx:Asympt_Vel_Plastron}) that in the limit of large gap length, $g\gg 1$, the normalised interfacial velocity follows the asymptotic hyperbolic trend
\begin{equation}\label{eq:uicScaling3}
\frac{u_{Ic}}{2(1-{\gamma}_{Ma})} \simeq \frac{1}{4-3\phi},
\end{equation} 
plotted with a black dashed line in figure~\ref{fig:uIc_with_g_and_phi}(\textit{b}). The asymptotic result (\ref{eq:uicScaling3}) is valid for any $\phi$. This result is consistent with (\ref{eq:uicScaling2}) and (\ref{eq:uIcasym1}).

\subsection{Predictions of the interfacial shear rate, effective slip length and drag reduction}
We now have two independent expressions relating the interfacial velocity $u_I$ and the Marangoni shear $\gamma_{Ma}$. The scaling (\ref{eq:scalinguIsurfactant}) was found based on near-interface surfactant dynamics, whereas (\ref{eq:uIscaled1}) was derived from a Stokes flow solution. Eliminating the interface velocity, we deduce a scaling expression for the average Marangoni shear rate,
%
\begin{equation}\label{eq:gammaMa}
    \gamma_{Ma} = a_1 k^*\mathcal{F}_0(g,\gf) \left(\frac{1}{Pe_I} + a_2 \frac{g^2 Bi}{1+\frac{Bi\,Pe}{\chi}\delta} + a_1 k^*\mathcal{F}_0(g,\gf) \right)^{-1},
\end{equation}
where $a_1$ and $a_2$ are the empirical parameters that were introduced in \S\ref{sec:scalingTheory}.
This predictive scaling depends only on the properties of the flow,  fluid and  surfactant through the non-dimensional numbers $k^*=k \,Ma$, $Pe_I$, $Bi$, $Pe$ and $\chi$, and on the two geometrical parameters $g$ and $\gf$. 
As noted earlier, it assumes a sufficiently small concentration  of surfactant and a small Reynolds number in the flow, and the diffusive boundary layer thickness $\delta$ depends only weakly on $\gamma_{Ma}$ following (\ref{eq:deltaPe12}), (\ref{eq:deltagPe12}) or (\ref{eq:deltagPe13}). The parameters $a_1$, $a_2$, as well as $\delta_{0,i}$ and $\delta_{1,i}$ (with $i=1$, 2 or 3 for the scaling predictions (\ref{eq:deltaPe12}), (\ref{eq:deltagPe12}) or (\ref{eq:deltagPe13}), respectively) for $\delta$, are determined empirically by fitting to our numerical simulations in \S\ref{sec:results}.



We can also compute a global effective slip length $\lambda_{e}$ as defined by \cite{Lauga_Stone_JFM_2003}, which corresponds to the \fernando{value $\lambda_{e}$ such that} an equivalent channel flow under the same pressure gradient, but with \fernando{a uniform Navier slip boundary condition $u(y=-1)=\lambda_{e}\left.\partial{u}/\partial{y}\right|_{y=-1}$ replaces the mixed conditions of the SHS at the bottom boundary}. We can show that the contribution of the effective slip length $\lambda_{e}$ is such that the total volume flux in the channel is the sum of the background Poiseuille volume flux, $Q_p=2/3$, and the volume flux of the deviating flow,
\begin{equation}\label{eq:lambdaeQ}
    Q = Q_p + Q_d = \frac{2}{3} + \frac{2{\lambda}_{e}}{{\lambda}_{e}+2},
\end{equation}
where the maximum value for the deviating flux is $Q_d\to 2$, as $\lambda_e\to \infty$.
The effective slip length as a function of the deviating flux is
\begin{equation}\label{eq:lambdaeQd}
    \lambda_{e} =  \frac{2{Q}_d}{2 - {Q}_d}.
\end{equation}
From equation (\ref{eq:Psidsimp}), the deviating streamfunction is
\begin{equation}
Q_d = {\Psi_d}(y=1)-\Psi_d(y=-1)=\frac{E}{2} - \left( -\frac{3E}{2}\right) = 2E,
\end{equation}
and substituting into (\ref{eq:lambdaeQd}) yields
\begin{equation}
\lambda_{e} = \frac{2E}{1-E}.
\end{equation}
Following from the linearity of the governing equations, and of the boundary conditions, $E$ also scales linearly with $(1-\gamma_{Ma})$. Accordingly, we can find the explicit dependence of $\lambda_{e}$ with the Marangoni shear rate $\gamma_{Ma}$,
\begin{equation}\label{eq:globaleffsliplength}
\lambda_{e} = \frac{2(1-\gamma_{Ma})E_0}{1-(1-\gamma_{Ma})E_0},
\end{equation}
where $E_0$ is the first coefficient of the vector $\bs{U}_n$ (see (\ref{eq:linsysmatrix})) in the surfactant-free case, \ie $E_0=E$ for $\gamma_{Ma}=0$, and $\gamma_{Ma}$ is expressed by (\ref{eq:gammaMa}). As expected, $0\leq \lambda_e < \infty$, since $0\leq\gamma_{Ma}\leq 1$ and $0\leq E_0 \leq 1$.

The corresponding drag reduction due to the presence of the SHS in our pressure-driven channel flows, inclusive of surfactant, can be computed as
\begin{equation}\label{eq:DRtemp}
    DR = 1 - \frac{C_f}{C_{f,p}} = 1 - \frac{\displaystyle\frac{\left<\hat{\tau}_s\right>}{\hat{\rho} (\hat{Q}/2)^2}}{\displaystyle\frac{\left<\hat{\tau}_s\right>_p}{\hat{\rho} (\hat{Q}_p/2)^2}}
\end{equation}
 where $C_f=\left<\hat{\tau}_s\right>/(\hat{\rho} (\hat{Q}/2)^2)$ is the laminar friction coefficient for a pressure-driven flow through a SHS channel with surfactants and $C_{f,p}=\left<\hat{\tau}_s\right>_p/(\hat{\rho} (\hat{Q}_p/2)^2)$ is the laminar friction coefficient for the equivalent Poiseuille channel flow driven with the same pressure gradient and for the same channel height. The quantities $\left<\tau_s\right>$ and $\left<\tau_s\right>_p$ are the surface stresses averaged along  both top and bottom surfaces for an SHS channel flow and a Poiseuille channel flow with the same geometry, respectively. Since, by construction, the pressure gradient is the same for the flow in the SHS channel and the Poiseuille channel flow, we  have $\left<\hat{\tau}_s\right>=\left<\hat{\tau}_s\right>_p$. Then, using (\ref{eq:lambdaeQ}) and (\ref{eq:lambdaeQd}) into~(\ref{eq:DRtemp}) we find
\begin{equation}\label{eq:DR}
    DR = 1 - \left(1+\frac{3Q_d}{2}\right)^{-2} = 1 - \left(1+\frac{3\lambda_e}{\lambda_e+2}\right)^{-2}.
\end{equation}
The maximum possible drag reduction is $DR\to 15/16$ as $\lambda_e\to \infty$.
We can compute $\lambda_e$ in (\ref{eq:DR}) using (\ref{eq:globaleffsliplength}) and (\ref{eq:gammaMa}). We also provide, as supplementary materials~\cite{githubroutine}, MATLAB routines computing $\lambda_e$, $DR$ and $\gamma_{Ma}$ for any specified flow-related, surfactant or geometrical parameters.

\section{Surfactant-laden numerical simulations}\label{sec:numericalsimulations}

To test the validity of our theoretical model and its predictions  for the surfactant-induced Marangoni shear $\gamma_{Ma}$ in (\ref{eq:gammaMa}) and for the effective slip length $\lambda_e$ in (\ref{eq:globaleffsliplength}), we performed 137 surfactant-laden numerical simulations of the full governing equations (\ref{eq:divu})--(\ref{eq:SurfNoFlux}). 

We varied  the nine dimensionless numbers independently over several
orders of magnitude to comprehensively explore the parameter space. As introduced in \S\ref{sec:probDesc}, these dimensionless groups are the Reynolds number $Re = \hat{\rho}\hat{h}\hat{U}/\hat{\mu}$,
the bulk and interface P\'eclet numbers $Pe = \hat{h}\hat{U}/\hat{D}$ and $Pe_I = \hat{h}\hat{U}/\hat{D}_I$, the Biot number  $Bi = \hat{\kappa}_d \hat{h}/\hat{U}$, the non-dimensional bulk concentration $k = \hat{\kappa}_a \hat{c}_0/\hat{\kappa}_d$, the surfactant adsorption--desorption kinetics number $\chi = \hat{\kappa}_d \hat{h} / (\hat{\kappa}_a \hat{\Gamma}_m)$, the Marangoni number $Ma = n_\sigma \hat{R}\hat{T} \hat{\Gamma}_m/(\hat{\mu} \hat{U})$, the gas fraction $\gf=\hat{g}/\hat{L}$ and the non-dimensional interfacial length $g=\hat{g}/\hat{h}$. The Frumkin interaction parameter, used in equation (\ref{eq:S}), is kept constant at  $A=-1$ for all our simulations. Since this parameter has a weak influence on the surfactant-induced Marangoni shear rate, we chose a value for $A$ corresponding to moderate attractive interactions between the adsorbed surfactant molecules. This value is close to the measured value for the common surfactant sodium dodecyl sulfate in de-ionised water: $A = -2.4$ \citep{Chang_Franses_1995,prosser01}.
The aim is also to obtain values for the  empirical parameters $a_1$, $a_2$ in  (\ref{eq:gammaMa}) and  $\delta_{0,i}$ and $\delta_{1,i}$ (with $i=1$, 2 or 3) in the uniform shear regime.

The model described by the dimensional form of equations (\ref{eq:divu}) to (\ref{eq:SurfNoFlux})  was implemented in COMSOL Multiphysics 5.2\textsuperscript{\tiny\textregistered} in two-dimensional finite-element numerical simulations. 
The SHS channel geometry shown in figure~\ref{fig:schematicSHSchannel}(\textit{a}) was used for the simulation domain, where the range of values for the gap length $\hat{g}$, the ridge length $\hat{l}$, the channel half-height $\hat{h}$ and the streamwise mean pressure drop per unit length $\hat{G}$ are presented in Supplementary Table S1.

When designing the mesh of the domain, we were particularly careful to ensure we could capture strong possible variations of some variables near the stagnation points at the beginning and end of the interface ($x=\pm g/2$), and in the vicinity of the interface. For each simulation, the maximum size of the mesh elements at the stagnation points, on the interface, and in the bulk, is detailed in Supplementary Table S1. Across all the simulations, the maximum density of elements close to the two stagnation points of the interface is 200 per micron, while the lowest density of elements at the middle of the interface is 20 per micron. 

To implement the model in COMSOL, we combine the Laminar Flow module with a Dilute Species Transport module for the transport equations in the bulk (\ref{eq:divu}--\ref{eq:C}). The equation for the transport of surfactant on the interface (\ref{eq:Gamma}) is implemented through a General Form Boundary PDE, with a source term corresponding to the Frumkin kinetics flux $S$ (\ref{eq:S}). This flux also serves to implement the condition for the continuity of the diffusive flux and the kinetics flux (\ref{eq:dCdy}) at the interface for the Dilute Species Transport module. The non-uniform distribution of surfactants at the interface yield Marangoni forces, which  modify the Laminar Flow module, as stated in (\ref{eq:MarangoniStress}), through a weak contribution at the interface coupled to a free-slip boundary condition, resulting in the required partial slip at the interface.

The flow is forced by a mean pressure drop per unit length, which is implemented through a Periodic Flow Condition between inlet and outlet \julien{following (\ref{eq:period_p})}, \francois{also enforcing velocity field periodicity between inlet and outlet}. A gauge for the pressure is imposed through a pressure point constraint at a corner of the domain. The initial guess for the velocity, for the stationary solver, is set to the reference Poiseuille profile $u_p=(1-y^2)/2$ in the entire chamber, corresponding to the stream-function (\ref{eq:Psip}). Periodic boundary conditions between inlet and outlet \julien{following (\ref{eq:period_c})} are also imposed in the Dilute Species Transport module for the bulk surfactant concentration $c$. 

\francois{To ensure the accuracy and stability of the numerical simulations}, we discretize the fluid flow with quadratic elements for the velocity field \francois{and} linear elements for the pressure field \francois{(Taylor-Hood elements)}, as well as quadratic elements for the concentration fields in the bulk and on the interface. We use the MUMPS solver of COMSOL to solve for the steady state of the system, with a relative tolerance of $10^{-5}$. \julien{All our 137 COMSOL numerical simulations were fully converged, satisfying this strict relative tolerance.}

\begin{table*}
    \centering
    \sisetup{
        table-number-alignment = center,
        table-figures-integer = 1,
        table-figures-decimal = 1,
        table-figures-exponent = 2,
        table-sign-exponent,
        range-phrase = --
    }
    \begin{adjustbox}{angle=90}
   \begin{tabular}{l
                    c
                    S[table-number-alignment = right,table-alignment = left]
                    S[table-number-alignment = right,table-alignment = left]
                    }
          \toprule
          {Parameter} & {Symbol} & {Minimum} & {Maximum} \\
          \midrule
          {Gas fraction} & $\phi=\hat{g}/\hat{L}$ & 1e-3 & 9.5e-1 \\
          {Length of the air--water interface} & $g=\hat{g}/\hat{h}$ & 1e-3 & 1e2 \\
          {Reynolds number} & $Re= \hat{\rho}\hat{h}\hat{U}/\hat{\mu}$ & 4e-4 & 1e5 \\
          {Bulk concentration} & $k= \hat{\kappa}_a \hat{c}_0/\hat{\kappa}_d$ & 1e-7 & 1e2 \\
          {Bulk P\'eclet number} & $Pe= \hat{h}\hat{U}/\hat{D}$ & 5e-6 & 2.5e7 \\
          {Bulk P\'eclet number (with $\hat{g}$)} & $Pe_g= \hat{g}\hat{U}/\hat{D}$ & 1e-6 & 1e6 \\
          {Interface P\'eclet number} & $Pe_I= \hat{h}\hat{U}/\hat{D}_I$ & 4 & 2e8 \\
          {Interface P\'eclet number (with $\hat{g}$)} & $\mathcal{F}_0 Pe_{I,g}= \mathcal{F}_0 \hat{g}\hat{U}/\hat{D}_I$ & 3.1e-4 & 2.5e5 \\
          {Biot number} & $Bi = \hat{\kappa}_d \hat{h}/\hat{U}$ & 1.2e-4 & 5e2 \\
          {Biot number} (with $\hat{g}$)  & $Bi_g = \hat{\kappa}_d \hat{g}/\hat{U}$ & 1.2e-5 & 2.5 \\
          {Kinetics number} & $\chi= \hat{\kappa}_d \hat{h} / (\hat{\kappa}_a \hat{\Gamma}_m)$ & 5e-3 & 5e3 \\
          {Kinetics number} (with $\hat{g}$) & $\chi_g= \hat{\kappa}_d \hat{g} / (\hat{\kappa}_a \hat{\Gamma}_m)$ & 2e-3 & 2e2 \\
          {Marangoni number} & $Ma= n_\sigma \hat{R}\hat{T} \hat{\Gamma}_m/(\hat{\mu} \hat{U})$ & 3 & 1.2e12 \\
          {Marangoni concentration} & $k^*= kMa$ & 3e-7 & 1.2e14 \\
          {Ratio of kinetics flux to} & & & \\
          {\quad advective flux at the interface} & $\mathcal{K}_{I,g}=Bi_g(1+k)/\mathcal{F}_0$ & 9.9e-4 & 3.2e3 \\
          {Ratio of diffusive flux to} & & & \\
          {\quad advective flux at the interface} & $\mathcal{D}_{I,g}=\chi_g(1+k)/(\mathcal{F}_0 Pe_g)^\frac{1}{2}$ & 4e-5 & 4.4e3 \\
          \bottomrule
    
    \end{tabular}
    \end{adjustbox}
\vspace{.3in}
        \caption{Range of values for all the non-dimensional parameters varied in the 137 finite element numerical simulations. Hatted quantities are dimensional. See also Supplementary Table S1 for the value of each parameter in each numerical simulation.}\label{tab:NDNminmax}
\end{table*}

The surfactant properties correspond to the well-characterized surfactant sodium dodecyl sulfate (SDS), which are well described by Frumkin kinetics \citep{prosser01}. The physical parameters were chosen in order to explore a large range of the key non-dimensional numbers. Variations by four to six orders of magnitude were explored, as summarized in table~\ref{tab:NDNminmax} in this section, as well as in figure~\ref{fig:parameterOverview} in appendix~\ref{apdx:keydimensionlessnumbers}.
\francois{In five simulations, we explored the limit of high Reynolds number with $Re\geq 1,000$, for which the flow should physically be at or above the transition to a turbulent regime. However, we imposed the flow to remain laminar in these simulations, since we are not interested in the effect of inertial instabilities or turbulence in this study.}
We will return to this point in \S\ref{sec:results}, when discussing results at large Reynolds numbers under laminar conditions. All other relevant physical and kinetics parameters of the 137 performed simulations are presented in Supplementary Table S1.

\section{Results and model performance}\label{sec:results}
\subsection{Effective slip length}\label{sec:EffectiveSlipLength}
\begin{figure}[t!]
\centering
\includegraphics[width=0.8\textwidth]{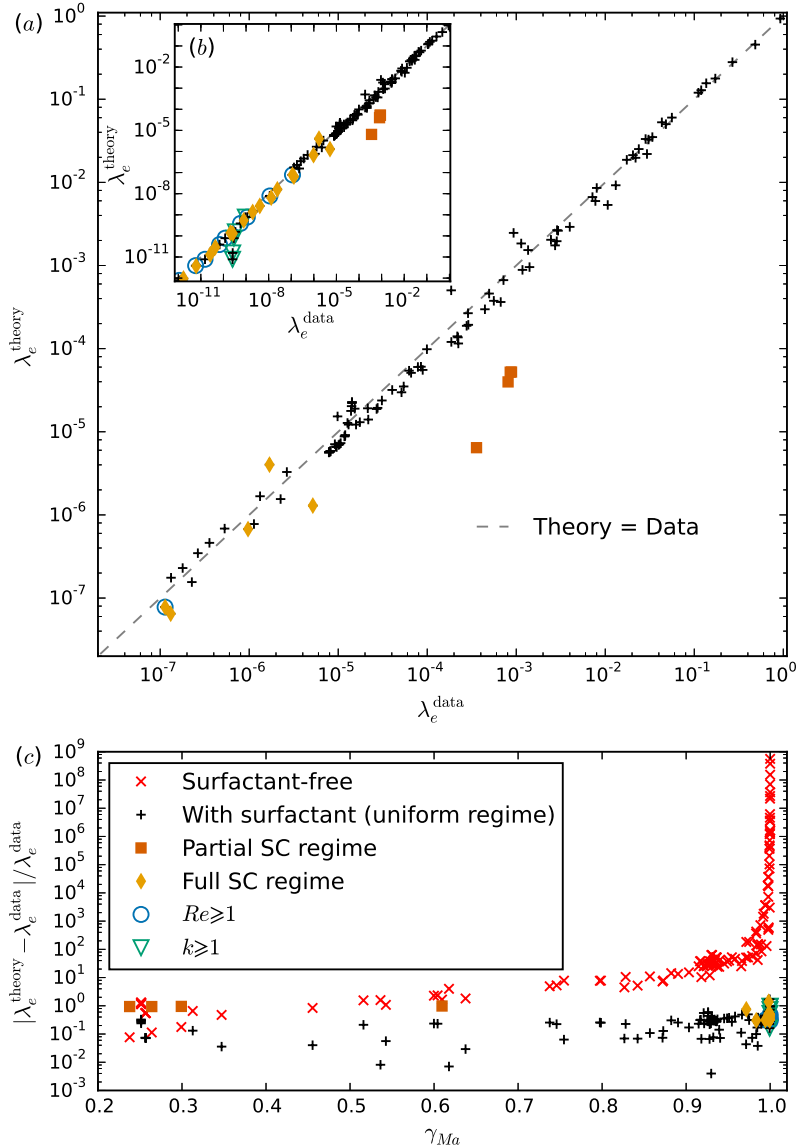}
 \caption{(\textit{a}) Comparison of the scaling predictions for the effective slip length $\lambda_{e}^\mathrm{theory}$, computed using (\ref{eq:globaleffsliplength}) and (\ref{eq:gammaMa}), with fitting parameters $a_1={2.30}$, $a_2 = {0.319}$, $\delta_{0,3}={1.68}$ and $\delta_{1,3}={0.0528}$, with the numerical results from our simulations  $\lambda_{e}^\mathrm{data}$, calculated from (\ref{eq:lambdaeQd}). Results are plotted on a log--log scale\francois{, with the grey dashed line showing equality between predictions and simulations}.  The predictions for the four data points in the partial stagnant cap (SC) regime, plotted with vermilion squares, underestimate the data owing to the strong non-uniformity of the interfacial shear rate profile. Nevertheless, the theory remains practically useful also for these cases, as it correctly predicts $\lambda_{e}\ll 1$. In the inset (\textit{b}), 
we plot 
an extended  range of $\lambda_{e}^\mathrm{data}$. In (\textit{c}), a linear--log plot shows the relative error between the data and the scaling predictions, as a function of the average interfacial shear rate. Red crosses show the error in the effective slip length when surfactants are neglected, such that  $\lambda_e$ is calculated using  (\ref{eq:globaleffsliplength}) with $\gamma_{Ma}=0$.
}
\label{fig:lambdaEffTheoryVSData}
\end{figure}

In \fig~\ref{fig:lambdaEffTheoryVSData}, we compare our scaling predictions for the effective slip length $\lambda_{e}^\mathrm{theory}$ with the numerical results  $\lambda_{e}^\mathrm{data}$. We compute $\lambda_{e}^\mathrm{theory}$ using (\ref{eq:globaleffsliplength}), where $\gamma_{Ma}$ follows (\ref{eq:gammaMa}) and the coefficients $E_0$ are computed by solving the linear problem (\ref{eq:linsysmatrix}) in the surfactant-free case for each couple of geometrical parameters $(g,\gf)$. The empirical parameters $a_1$, $a_2$ in (\ref{eq:gammaMa}) and $\delta_{0,i}$, $\delta_{1,i}$, with $i=1$, 2 or 3 for $\delta$ (see equations~(\ref{eq:deltaPe12})--(\ref{eq:deltagPe13})) can be determined using a least-squares fitting approach and the Trust Region Reflective algorithm, as implemented in the package \verb|optimize.least_squares| of Scipy \citep{Scipy}. 

First, we determine  $\delta_{1,3}$ in $\delta$ by fitting a measure of the characteristic diffusive boundary layer thickness in our numerical simulations, calculated using (\ref{eq:SurfactantFluxScalingKineticsBulk2}), with the scaling model given in (\ref{eq:deltagPe13}). We have only used the \Leveque scaling (\ref{eq:deltagPe13}) for $\delta$ in (\ref{eq:gammaMa}). In our numerical simulations, the diffusive boundary layer mostly follows the \Leveque regime, which assumes a background linear shear flow, since the slip velocity $u_I$ is small. Moreover, as also noted earlier, the scaling model (\ref{eq:gammaMa}) for $\gamma_{Ma}$ depends weakly on $\delta$. Hence, the choice of scaling for  $\delta$, which can vary between  (\ref{eq:deltaPe12}), (\ref{eq:deltagPe12}) or (\ref{eq:deltagPe13}) depending on the geometry and the slip, does not appear to be critical. The fit gives 
\begin{equation}
\delta_{1,3}={0.0528},
\end{equation}
%
from the minimization of the sum of the squares of the relative distance of theory from data, \ie $(\delta^\mathrm{theory}-\delta^\mathrm{data})^2/(\delta^\mathrm{data})^2$.
This prior independent determination of $\delta_{1,3}$ reduces the number of fitting parameters to three in (\ref{eq:gammaMa}): $a_1$, $a_2$ and $\delta_{0,3}$. This ensures a more accurate and robust fit for $\lambda_{e}^\mathrm{theory}$, less sensitive on the actual fitting technique used.

Then, using $\delta_{1,3}=0.0528$ in (\ref{eq:gammaMa}), we fit the effective slip length $\lambda_{e}^\mathrm{theory}$ given by  (\ref{eq:globaleffsliplength}) to  $\lambda_{e}^\mathrm{data}$ computed via the deviating flux $Q_{d}$ using (\ref{eq:lambdaeQd}). Incidentally, computing $\lambda_{e}^\mathrm{data}$ using (\ref{eq:lambdaeQd}) gives an accurate and robust estimation of the effective slip length in our numerical simulations, as it relies solely on the integral quantity $Q_{d}=Q-Q_p$ (see \ref{eq:lambdaeQ}). Minimising the sum of the squares of the absolute distance between $\lambda_{e}^\mathrm{theory}$ and $\lambda_{e}^\mathrm{data}$, 
we obtain 
\begin{equation}
a_1={2.30},\ a_2 = {0.319} \ \textrm{ and }\ \delta_{0,3}={1.68}. 
\end{equation}
 

As we can see in \fig~\ref{fig:lambdaEffTheoryVSData}(\textit{a},\textit{b}), the scaling predictions for $\lambda_{e}^\mathrm{theory}$, using the values for $a_1$, $a_2$, $\delta_{0,3}$ and $\delta_{1,3}$ stated in the previous paragraph, show an excellent agreement with $\lambda_{e}^\mathrm{data}$ over a very large range:  $\num{e-12}\lesssim \lambda_e \lesssim 1$. 

The nine data points at non-negligible Reynolds numbers, $1\leq Re\leq \num{e5}$ (identified with blue circles in \fig~\ref{fig:lambdaEffTheoryVSData}(\textit{a},\textit{b})), also exhibit good agreement despite violating the low Reynolds number assumption made in our flow model (see \S\ref{sec:StokesFlowModel}). As explained previously in \S\ref{sec:numericalsimulations}, although the full steady nonlinear Navier--Stokes equation (\ref{eq:u}) was used in the simulations, the flow remained in the laminar regime for all Reynolds numbers tested. 

At large non-dimensional background concentrations, $1 \leq k \leq 100$ (identified with green triangles in \fig~\ref{fig:lambdaEffTheoryVSData}(\textit{b})), the scaling predictions underestimate slightly the slip length. This is due to the fact that the model assumes a low concentration of surfactant. However, the model still provides a practically useful 
prediction of the boundary condition at the interface, which can be effectively considered as no-slip for all our simulations with $k\geq 1$.
We also find that the maximum boundary layer thickness is $\delta =1.20$, which suggests that our scaling prediction is accurate even if the diffusive boundary layer is vertically confined.

We  indicate in \fig~\ref{fig:lambdaEffTheoryVSData} (as well as in \fig{s}~\ref{fig:DRTheoryVSData} and \ref{fig:gammaMaTheoryVSData}) data where the interface properties are strongly nonuniform, which are labeled by vermilion squares and orange diamonds. 
Qualitatively similar interface non-uniformities have been studied extensively in the context of air bubbles rising in surfactant-contaminated water \cite[\eg][]{bond28,frumkin47,levich62,he91}, where they correspond to the `stagnant cap regime'. In this regime, an upstream part of the interface has a negligible surfactant gradient and can be considered as shear-free ($\gamma_{Ma}\to 0$), whilst the rest of the interface downstream has a large Marangoni shear ($\gamma_{Ma}\to 1$), leading to a no-slip condition over a portion of the bubble known as the `stagnant cap' (hereafter designated as SC). 
In the SC regime, advection of surfactant along the interface dominates relative to surfactant transport between the interface and the bulk. This makes possible highly non-uniform interfacial concentrations.
Since transport between the interface and the bulk is mediated by both the diffusive boundary layer flux and the surfactant kinetics, the SC regime requires that advection along the interface must be large compared to either diffusive or kinetic fluxes (or both).

{We briefly summarize here the bubble-flow analysis of \cite{Palaparthi2006-di}, and translate it to SHS flow. 
For a bubble, the SC regime 
is found when the characteristic interfacial \Peclet number 
is large, and either the adsorption--desorption kinetics flux $S$ 
is small, or the diffusive flux through the boundary layer is small 
compared with the interfacial advective flux. Denoting with a superscript `bubble' the results of \cite{Palaparthi2006-di}, they showed that this implies $Pe_{I}^\text{bubble} \gg 1$, and $\mathcal{K}_{I} = Bi^\text{bubble}(1+k)\ll 1$ or $\mathcal{D}_{I}=\chi^\text{bubble}(1+k)/(Pe^\text{bubble})^{1/2}\ll 1$. For a bubble, the characteristic length and velocity scales are the bubble radius and interfacial velocity in the surfactant-free case. 
In order to translate these canonical bubble results to SHSs, note that the bubble radius is analogous to the grating length $\hat{g}$. 
For the SHS, the characteristic velocity scale for these non-dimensional numbers is the mid-gap interfacial velocity in the surfactant-free case, namely $\hat{u}_{Ic}(\gamma_{Ma}=0)$, which differs from the bulk characteristic velocity, such that $\hat{u}_{Ic}(\gamma_{Ma}=0)=2 \mathcal{F}_0 \hat{U}$, according to~(\ref{eq:uIscaled1}). 
This contrasts slightly with contaminated air bubbles in water, where the characteristic interfacial velocity in the surfactant-free case scales as the far-field bulk velocity, owing to the absence of rigid no-slip walls. As shown in \fig~\ref{fig:uIc_with_g_and_phi} and explained in detail in appendix~\ref{apdx:Asympt_Vel_Plastron}, we have $\mathcal{F}_0 \sim 1$ for $g\gtrsim 1$ (as for bubbles) and $\mathcal{F}_0 \sim g$ for $g\lesssim 1$.

Therefore, using our dimensionless group definitions of \S\ref{sec:probDesc}, and using a `$g$' subscript to characterize dimensionless groups where we use the lengthscale $\hat{g}$, rather than $\hat{h}$, we have $Pe_{I}^\text{bubble} \mapsto \mathcal{F}_0 g Pe_{I} = \mathcal{F}_0 Pe_{I,g}$ and $\mathcal{K}_{I} = Bi^\text{bubble}(1+k) \mapsto \mathcal{K}_{I,g} = Bi_g(1+k)/\mathcal{F}_0$, as well as $\mathcal{D}_{I}=\chi^\text{bubble}(1+k)/(Pe^\text{bubble})^{1/2} \mapsto  \mathcal{D}_{I,g}=\chi_g(1+k)/(\mathcal{F}_0 Pe_g)^{1/2}$.
}
%
%
The ranges spanned by the quantities $\mathcal{F}_0 Pe_{I,g}$, $\mathcal{K}_{I,g}$ and $\mathcal{D}_{I,g}$ 
are reported in table~\ref{tab:NDNminmax}. 

The distinction between the partial SC regime, where the SC fills only part of the interface, and the full SC regime, where the SC fills all the interface, is 
revealed by an inspection of the  shear rate profiles along the interface (not shown here). In the partial SC regime, the shear rate increases abruptly from negligible values to $\gamma_{Ma}\sim 1$ at a particular location along the interface.
In the partial SC regime, the non-dimensional numbers in our simulations range approximately: $\num{2.5e3}\leq \mathcal{F}_0 Pe_{I,g}\leq \num{2.5e5}$, $\num{9.9e-4}\leq \mathcal{K}_{I,g}\leq 0.4$ and $0.04\leq \mathcal{D}_{I,g}\leq 0.4$ (see also \fig~\ref{fig:parameterOverview}, appendix~\ref{apdx:keydimensionlessnumbers}, for the variations of these numbers across all our numerical simulations and for the different regimes, as well as Supplementary Table S1 for the value of each parameter for each simulation). In the full SC regime, the non-dimensional numbers range approximately: $52\leq \mathcal{F}_0 Pe_{I,g}\leq \num{2.5e4}$, $\num{2e-2}\leq \mathcal{K}_{I,g}\leq 50$ and $\num{4.0e-5}\leq \mathcal{D}_{I,g}\leq 1.3$. The interfacial \Peclet number is mostly higher in the partial SC regime than in the full SC regime, which is intuitively expected.
We can see in \fig~\ref{fig:lambdaEffTheoryVSData}(\textit{a}) that the four data points in the partial SC regime (plotted with vermilion squares) are the only data points where the scaling predictions significantly underestimates the effective slip length with $\lambda_{e}^\mathrm{theory}\leq {\num{5.2e-5}}$, whereas $\lambda_{e}^\mathrm{data}\geq \num{3.5e-4}$. This discrepancy is due to the strong non-uniformity of the shear rate profile in the SC regime, not taken into account by our scaling model which is based on the assumption that the shear rate is approximately uniform along the interface (see (\ref{eq:ScalingGammaMa})).
The predictions $\lambda_{e}^\mathrm{theory}$ in the full SC regime, plotted with orange diamonds, are in reasonable agreement with the data $\lambda_{e}^\mathrm{data}$. We can see that $\lambda_{e}^\mathrm{theory}$ underestimates slightly the data, although by less than one order of magnitude for all our results in the full SC regime, with ${0.25}\leq \lambda_{e}^\mathrm{theory}/\lambda_{e}^\mathrm{data}\leq {2.4}$.

The data plotted with black pluses in \fig~\ref{fig:lambdaEffTheoryVSData}, \ie not in the SC regime, are in a state analogous to the `uniformly retarded regime' described by \cite{Palaparthi2006-di} in their study of air bubbles rising in contaminated water, where they make the case that this regime exists for $\mathcal{K}_{I,g}\sim 1$ and $\mathcal{D}_{I,g}\sim 1$. However, in our simulations we find that the  interfacial shear rate is  in the `uniform' regime, and thus satisfies our modelling assumption, over a range of $\mathcal{K}_{I,g}$ and $\mathcal{D}_{I,g}$ that spans several orders of magnitude, implying that the vast majority of the simulations satisfy our modelling assumptions. More specifically, we find that simulations in the `uniformly retarded regime' have parameters that satisfy approximately $\num{2.8e-3}\leq \mathcal{D}_{I,g} \leq\num{4.4e3}$ and $\num{1.9e-2}\leq \mathcal{K}_{I,g} \leq \num{3.2e3}$.  This is most likely due to the fact that some of our simulations are in an intermediate or transition regime between the SC regime and the uniformly retarded regime, and for which $\lambda_e$ still follows our scaling prediction, though perhaps with slightly more scatter, as shown by some of the black pluses in \fig~\ref{fig:lambdaEffTheoryVSData}.

In \fig~\ref{fig:lambdaEffTheoryVSData}(\textit{c}), we show the relative error between the scaling predictions $\lambda_{e}^\mathrm{theory}$ and the numerical results $\lambda_{e}^\mathrm{data}$ for the effective slip length. The error remains relatively small across all values of the average interfacial shear rate $\gamma_{Ma}$. It is less than approximately {33\%} for $\gamma_{Ma}\leq 0.7$, except for the four simulations in the partial SC regime  plotted with vermilion squares. The relative error is less than {1.7}  for $0.7\leq \gamma_{Ma}\leq 1$.

For comparison, we also show with red crosses in \fig~\ref{fig:lambdaEffTheoryVSData}(\textit{c}) the prediction from a surfactant-free model, which is obtained using (\ref{eq:globaleffsliplength}) with $\gamma_{Ma}=0$.
Our model provides consistently better predictions than the one that neglects surfactant effect. 
In particular, the error made by neglecting surfactant effects becomes very large when the interfacial shear rate increases towards the Poiseuille value $\gamma_p=1$. At low shear rate, $\gamma_{Ma}\leq 0.3$ we can see that the two models have comparable (small) relative errors.

Overall, we find that our scaling model for $\lambda_e$ provides excellent quantitative predictions across a large range of non-dimensional numbers, beyond the strict range of validity based on our modelling assumptions. Although our model predictions can underestimate the slip length in some cases  (at large concentrations, and in the stagnant cap regime), our model remains practically useful as both theory and simulation yield negligible slip in those instances.


\subsection{Drag reduction}

\begin{figure}[t!]
\centering
\includegraphics[width=0.8\textwidth]{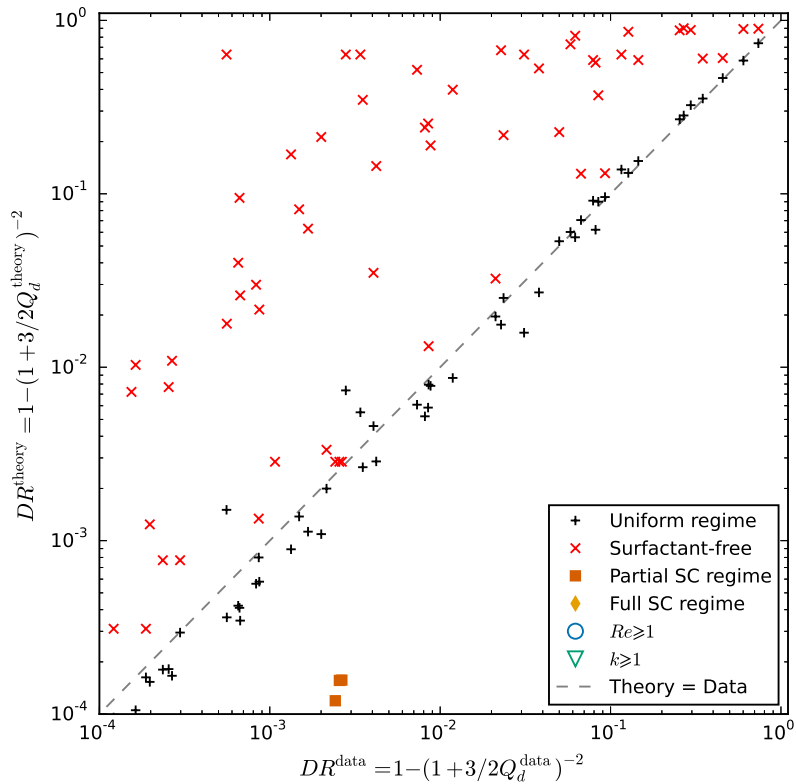}
\caption{
Comparison of the scaling predictions for the drag reduction $DR^\mathrm{theory}$, computed using \eqref{eq:DR}, with the numerical results from our simulations $DR^\mathrm{data}$. 
We also show with red crosses the drag reduction $DR^\mathrm{theory}$ estimated using a model neglecting surfactant effect. Note that we plot in this graph only data for $DR\geq \num{e-4}$, to show more clearly our results in a range useful to applications. \francois{Equality between data and theory falls on the grey dashed line.}
}
\label{fig:DRTheoryVSData}
\end{figure}

We compare the drag reduction predicted by our theory ($DR^\mathrm{theory}$)  with the numerical results from our simulations ($DR^\mathrm{data}$), as shown in \fig~\ref{fig:DRTheoryVSData}. 
The value of $DR^\mathrm{theory}$ is obtained from (\ref{eq:DR}), where the corresponding values of $\lambda_{e}^\mathrm{theory}$ are shown earlier in \fig~\ref{fig:lambdaEffTheoryVSData}. Similarly, $DR^\mathrm{data}$, is calculated using $\lambda_{e}^\mathrm{data}$, whose values are also shown in \fig~\ref{fig:lambdaEffTheoryVSData}.
%
Using a log--log scale, we only plot  data for $DR\geq \num{e-4}$, which correspond to the more meaningful range for practical applications. The predictions from our scaling model are in very good agreement with the numerical results. Data at even lower drag reductions (not shown here) still exhibit a very good agreement with our theoretical prediction.

In \fig~\ref{fig:DRTheoryVSData}, we also plot, using red crosses, the drag reduction computed using a surfactant-free model. This is obtained by substituting the values for the surfactant-free $\lambda_{e}^\mathrm{theory}$ (plotted with red crosses in \fig~\ref{fig:lambdaEffTheoryVSData}\textit{c}) into (\ref{eq:DR}). As may be expected, the surfactant-free theory almost always incorrectly predicts a larger drag reduction, with values often more than an order of magnitude larger that the actual ones. This clearly shows that the drag reduction potential of SHSs can be significantly overestimated in conditions where surfactants are important. This is consistent with the findings of \cite{peaudecerf17}, who showed that, for SHSs with rectangular longitudinal gratings, surfactant effects become important at very low concentrations, similar to background levels found in the environment.
As may be expected, the few surfactant-free predictions in \fig~\ref{fig:DRTheoryVSData} that show better agreement with the numerical simulations correspond to lower values of $\gamma_{Ma}$, when the surfactant-free predictions converge towards our model predictions (see \fig~\ref{fig:lambdaEffTheoryVSData}\textit{c}).

\subsection{Interfacial shear rate}\label{sec:interfacialshear}


\begin{figure}[t!]
\centering
\includegraphics[width=0.8\textwidth]{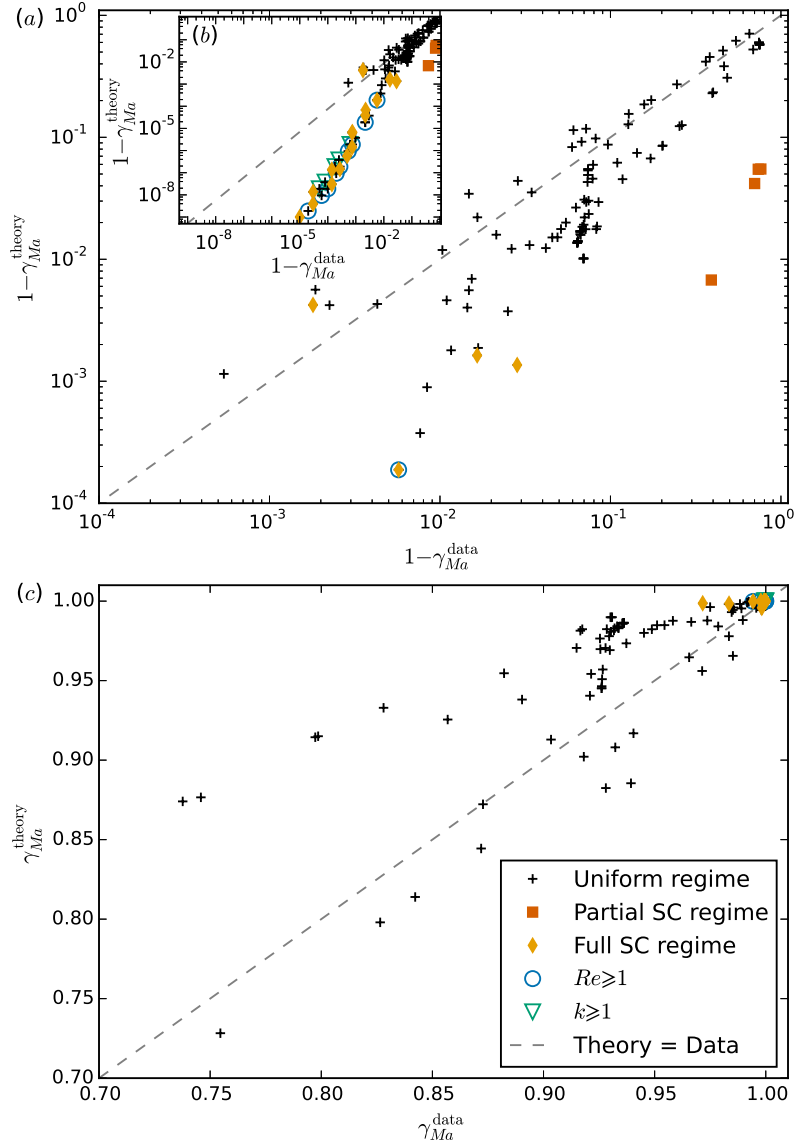}
\caption{
Comparison of the scaling predictions for the average interfacial shear rate $\gamma_{Ma}^\mathrm{theory}$, computed using (\ref{eq:gammaMa}), with the numerical results from our simulations  $\gamma_{Ma}^\mathrm{data}$, calculated by averaging the shear rate along the interface. The scaling prediction use empirical parameters:  $a_1={2.30}$ and $a_2 = {0.319}$ for $\gamma_{Ma}$, and $\delta_{0,3}={1.68}$ and $\delta_{1,3}={0.0528}$ for $\delta$ (see (\ref{eq:deltagPe13})), computed from the fit of $\lambda_e$ (see \S\ref{sec:EffectiveSlipLength}). In (\textit{a}), we plot using a log--log scale $(1-\gamma_{Ma})$ to reveal the behaviour at large shear rate, when $\gamma_{Ma}^\mathrm{data}\to \gamma_p =1$. The predictions $\gamma_{Ma}^\mathrm{theory}$ for the four data points in the partial stagnant cap (SC) regime, plotted with vermilion squares, overestimate the data owing to the strong non-uniformity of the interfacial shear rate profile. 
In (\textit{b}), we plot 
$(1-\gamma_{Ma})$ over a larger range revealing the error related to the singularity at the stagnation points. In (\textit{c}), we plot 
$\gamma_{Ma}$  to show more clearly the behaviour at intermediate shear rate, $0.7\leq \gamma_{Ma}\leq 1$. \francois{In all plots, equality between data and theory would fall on the grey dashed line.}
}
\label{fig:gammaMaTheoryVSData}
\end{figure}

We compare in \fig~\ref{fig:gammaMaTheoryVSData} the numerical results for the average interfacial shear rate $\gamma_{Ma}^\mathrm{data}$ with the theoretical predictions, $\gamma_{Ma}^\mathrm{theory}$ computed using (\ref{eq:gammaMa}) using the four empirical parameters optimized for $\lambda_e$ in \S\ref{sec:EffectiveSlipLength}: $a_1={2.30}$ and $a_2 ={ 0.319}$ for $\gamma_{Ma}$, and $\delta_{0,3}={1.68}$ and $\delta_{1,3}={0.0528}$ for $\delta$ based on (\ref{eq:deltagPe13}). The numerical results for $\gamma_{Ma}^\mathrm{data}$ have been computed by taking the spatial average of the interfacial shear rate in the interior of the interface $-g/2\leq x\leq g/2$.

In \fig~\ref{fig:gammaMaTheoryVSData}(\textit{a}), we show $(1-\gamma_{Ma})$ in a log--log plot to focus on the no-slip limit $\gamma_{Ma}\to 1$. Over the limited range shown on this graph, we find good agreement between our scaling predictions 
and the data 
for all our numerical simulations where the interfacial shear rate is found approximately uniform along the interface (see the uniform regime, plotted with black plusses). 
Similar to $\lambda_e$ shown in \fig~\ref{fig:lambdaEffTheoryVSData}, we can see that the four data points in the partial SC regime (vermilion squares) with $(1-\gamma_{Ma}^\mathrm{data})\geq 0.4$ are the only ones where the predictions 
underestimate the data. 
As discussed earlier, this is due to the strong non-uniformity of the shear rate profile in the SC regime, in contradiction with the uniform assumption made in our model (see (\ref{eq:ScalingGammaMa})). Nevertheless, the theory remains practically useful, as both model and simulation yield shear that is essentially indistinguishable from that of a no-slip boundary.

In \fig~\ref{fig:gammaMaTheoryVSData}(\textit{b}) (which is the inset of \fig~\ref{fig:gammaMaTheoryVSData}\textit{a}), we plot $(1-\gamma_{Ma}^\mathrm{data})$ over the full range of values tested. As the average shear rate tends to the maximum Poiseuille value, $\gamma_{Ma}^\mathrm{data}\to \gamma_p=1$ or equivalently $(1-\gamma_{Ma}^\mathrm{data})\to 0$, $(1-\gamma_{Ma}^\mathrm{theory})$ underestimates the data. The difference becomes significant for $(1-\gamma_{Ma}^\mathrm{theory})\lesssim \num{e-3}$. This is due to the singularity at the two stagnation points and the difficulty associated with resolving it numerically. The shear rate exhibits extreme variations very close to the stagnation points, whilst the shear rate remains flat in the interior of the interface with values very close to the Poiseuille shear rate. We note however that the effect of the singularity appears only in the limit $\gamma_{Ma}^\mathrm{data}\to \gamma_p=1$, at values practically equivalent to a no-slip boundary condition at the interface.

This can also be seen in \fig~\ref{fig:gammaMaTheoryVSData}(\textit{c}), where we plot $\gamma_{Ma}$ directly, for $\gamma_{Ma}\geq 0.7$. The scaling predictions consistently predict a no-slip boundary condition $\gamma_{Ma}^\mathrm{theory}\to 1$, as $\gamma_{Ma}^\mathrm{data}\to 1$. This shows that the actual error between $\gamma_{Ma}^\mathrm{theory}$ and $\gamma_{Ma}^\mathrm{data}$ is actually very small in this limit, where we find the simulations at large Reynolds numbers $Re\geq 1$ (blue circles), large non-dimensional concentrations $k\geq 1$ (green triangles) and in the full SC regime (orange diamonds). Predictions at intermediate values (shown by plusses in \fig~\ref{fig:gammaMaTheoryVSData}\textit{c}), for $0.7\leq \gamma_{Ma}^\mathrm{theory} \leq 1$, show a good agreement with $\gamma_{Ma}^\mathrm{data}$ although with a slight overestimation.

Therefore, our scaling model also provides reasonable predictions across the whole range of interfacial shear rate values, even though the model has been fitted for $\lambda_e$ and not for $\gamma_{Ma}$. An agreement is found from intermediate to large  values, provided the interface is not in a partial SC regime. Our scaling model remains accurate across a broad range of non-dimensional numbers (see table~\ref{tab:NDNminmax} and \fig~\ref{fig:parameterOverview}, appendix~\ref{apdx:keydimensionlessnumbers}) and in the full SC regime.

\section{Discussion}\label{sec:disc}
\subsection{\paolo{Verifying the validity of our main assumptions}}
The first key assumption  in our scaling model is that the non-dimensional interfacial surfactant concentration $\Gamma$ is sufficiently small so that the adsorption--desorption kinetics flux $S$ in (\ref{eq:S}) and the coupling condition (\ref{eq:MarangoniStress}) between the viscous stress and the surfactant-induced Marangoni stress can be linearised (see \S\ref{sec:kStar}). To test the validity of the assumption $\Gamma\ll 1$, at least a posteriori, 
we can note that it implies $\Gamma\sim k \ll 1$, which results from applying (\ref{eq:Slin}) at $S=0$ along the interface.
As mentioned before, we expect that $k$ should remain low in many applications where surfactants are not artificially added. \cite{peaudecerf17} estimated typical ranges of $k$, depending on whether one considers `weak' or `strong' surfactant. 
\cite{peaudecerf17} calculated that, for `weak' types of surfactants, the non-dimensional concentration range is $\num{e-9}\lesssim k\lesssim \num{e-2}$, which supports our hypothesis. Note that the upper bound of this range is given at the critical micellar concentration for the bulk concentration $\hat{c}_0$, implying that the worst-case scenario corresponds to water that is saturated with surfactant. Only for `strong' types of surfactant they indicated that the $k\ll 1$ assumption could potentially be invalid, since $\num{e-6}\lesssim k\lesssim \num{e3}$. Strong surfactants are likely to be found only in applications where they have been artificially added. Nevertheless, the model presented here performed well even at large $k$, as seen for example in \fig~\ref{fig:lambdaEffTheoryVSData}.

The second key assumption made in our scaling model is that the surfactant-induced Marangoni shear rate along the interface is approximately uniform. This is related to having a uniform concentration gradient, following the linearised coupling condition (\ref{eq:MarangoniStresslin}). From the broad range of parameters tested, see table~\ref{tab:NDNminmax} and \fig~\ref{fig:parameterOverview}, appendix~\ref{apdx:keydimensionlessnumbers}, we find that this assumption is  invalid only in the partial stagnant cap (SC) regime, where the concentration gradient presents an abrupt increase at some point along the interface, separating the no-shear and no-slip regions. As we saw in \fig~\ref{fig:lambdaEffTheoryVSData}, in the partial SC regime (see vermilion squares) our model underestimates the slip length. However, it is noteworthy that our scaling model provides reasonably accurate predictions for the full SC regime, where the no-slip region spans the whole interface. Furthermore, our scaling model remains practically useful in both the partial and full SC regimes, since it correctly predicts an essentially negligible effective slip length.

If wishing to strictly determine whether our model applies, we must therefore distinguish the parameter ranges between the full and partial SC regimes. As explained in \S\ref{sec:EffectiveSlipLength}, the SC regime exists when the  \Peclet number at the interface, $\mathcal{F}_0 Pe_{I,g}$, is large and either $\mathcal{D}_{I,g}$ or $\mathcal{K}_{I,g}$ are small.  From our simulations, we cannot find any clear distinction between the partial and full SC regimes based only on $\mathcal{D}_{I,g}$ or $\mathcal{K}_{I,g}$. However, we noted already that the partial SC regime was generally found at larger \Peclet numbers, for $\mathcal{F}_0 Pe_{I,g}\gtrsim \num{e3}$, whilst the full SC regime was found for $1\ll \mathcal{F}_0 Pe_{I,g}\lesssim \num{e4}$. This is physically intuitive as increasing the external flow velocity would eventually overcome the Marangoni stress at the interface. This would lead to a compression of the finite amount of surfactant adsorbed onto the interface towards the downstream end, thereby freeing the upstream part of the interface from any shear.

Since $\mathcal{F}_0 Pe_{I,g}\propto \mathcal{F}_0 \hat{U}$, $\mathcal{K}_{I,g}\propto 1/(\mathcal{F}_0 \hat{U})$ and $\mathcal{D}_{I,g}\propto 1/(\mathcal{F}_0 \hat{U})^{1/2}$, we expect to find the partial SC regime 
in applications where the  characteristic velocity near the interface $\mathcal{F}_0 \hat{U}$ is  large. We emphasize again that the characteristic velocity in these dimensionless numbers is the local characteristic velocity near the interface, $\mathcal{F}_0 \hat{U}$, where the bulk velocity $\hat{U}$ is modulated by  the geometrical function $0\leq \mathcal{F}_0\leq 1$, which scales as $\mathcal{F}_0\sim  g$ for $g \lesssim 1$, otherwise $\mathcal{F}_0\sim  1$.
Hence, our model is valid for applications at sufficiently low $\hat{U}$ or if $g$ is sufficiently small such that the SHS is away from the partial SC regime. Microfluidic applications, such as lab-on-a-chip systems or micro-cooling, where $\hat{U}$ is 
\paolo{small} 
would be typical applications for our model. 
For instance, we can consider a typical micro-fluidic channel  with  $\Hat{h}=50$ microns, a flow of water with characteristic speed ranging $0.1$ to $\SI{10}{mm \per s}$, and SHS gratings of length $\Hat{g}=\SI{1}{mm}$ with gas fraction $\gf\approx 0.95$. If surfactants similar to sodium dodecyl sulfate are present at a concentration of approximately $\SI{e-3}{mM}$ (equivalent to traces naturally present in the water), then we obtain: $800\leq \mathcal{F}_0 Pe_{I,g}\leq \num{8e4}$, $12\leq \mathcal{K}_{I,g}\leq \num{120}$, $0.7\leq \mathcal{D}_{I,g}\leq \num{7}$, and $k=\num{e-3}$. This shows that for this geometry with this range of flow speeds, the SHS would be in the uniform regime, far from the stagnant cap regime, such that our model would predict accurate estimates of the impact of surfactant on the slip length, drag reduction and  average Marangoni shear rate.

\subsection{\paolo{Comparison to experimental studies of surfactant effects}}
\julien{Another application of our model is to analyze experimental studies reporting  degradation of the performance  of SHSs where surfactant contamination could be the cause. For example, two recent studies by \cite{peaudecerf17} and \cite{Song2018-uw}  identified surfactant as the cause for the reduced or negligible slip measured near SHSs in laminar channel flows. As we discuss in detail in appendix~\ref{apdx:discussion-PNAS-Song}, the main difficulty in applying our model to predict the reduction in slip in their experiments is the absence of information regarding the surfactant properties and concentration. This is due to the fact that the surfactants were not introduced artificially, but were present as unwanted and unknown contaminants in their experiments.
}

\julien{Nevertheless, we can use our model to analyze \emph{a posteriori} the impact of surfactant in the studies of \cite{peaudecerf17} and \cite{Song2018-uw}. Assuming different possible surfactant types, we find that our theoretical model can predict physically sensible concentrations $\hat{c}_0$, which would lead to the reduced slip measured in  their experiments. As detailed in appendix~\ref{apdx:discussion-PNAS-Song},  our model predicts that for instance a `strong' surfactant \cite[see][]{peaudecerf17} would only require minute traces, far below typical environmental concentrations, to reduce the slip velocity $u_I$ as measured by \cite{peaudecerf17} and \cite{Song2018-uw}. A weak surfactant, \eg SDS, would require large $\hat{c}_0$ close to the critical micellar concentration, whilst an intermediate surfactant (see appendix~\ref{apdx:discussion-PNAS-Song}) would require small $\hat{c}_0$ at or below typical environmental conditions. Hence, our model predictions are consistent with the experimental results of \cite{peaudecerf17} and \cite{Song2018-uw} attributing their reduced performance to surfactant contaminant traces. Our model can also provide a rational \emph{a posteriori} explanation for other   experimental and field measurements that have reported poor SHS drag reduction performance, in contradiction with surfactant-free theoretical or numerical predictions. Therefore, our model could help design future SHSs to mitigate or avoid surfactant effect \emph{a priori}, for instance by identifying the optimal geometry and flow conditions for a given surfactant contaminant.
}

\subsection{\paolo{Analytical limits for slip and drag}}
\paolo{It can be instructive and useful to examine practically relevant limits where our results simplify. We discuss effects of key dimensionless groups, and derive expressions in the limits of insoluble surfactant, and of very long gratings. In the latter case, it is possible to immediately predict the drag reduction without the need for solving the full Stokes flow problem. In any other case, we recommend using the MATLAB codes provided as supplementary materials~\cite{githubroutine}.}

\paolo{To model insoluble surfactant, consider the interfacial advection--diffusion equation (\ref{eq:GammaStar}), setting the kinetics term on the right-hand side to zero. Integrating from the upstream stagnation point $x=-g/2$ to $x_0$, and dividing through by $\left. \Gamma^* \right|_{x_0}$, we obtain
\begin{equation}\label{eq:uI_ins}
    \left. u_I\right|_{x_0,\text{ins}} =  \frac{\gamma_{Ma}}{\left.\Gamma^* \right|_{x_0} \, Pe_I}\ .
\end{equation}}
%
%
%
%
%
\paolo{Dividing by $\gamma_{Ma}$ we find the plastron slip length, in the insoluble limit
\begin{equation}
   \left. \lambda_{x_0} \right|_\text{ins} = \frac{1}{\left.\Gamma^*\right|_{x_0}\, Pe_I} \simeq \frac{1}{Ma_\text{ins}},
\end{equation}
Where we assume that $ \left.\Gamma \right|_{x_0} \simeq \Gamma_s$, where $\Gamma_s$ is the (uniform) interfacial concentration found in static conditions, and $ Ma_\text{ins}$ is a Marangoni number for insoluble surfactant, namely
\begin{equation}
  Ma_\text{ins} =  \Gamma^*_{s} Pe_I = \Gamma_s Ma\, Pe_I = \frac{\hat{\Gamma}_s}{\hat{\Gamma}_m} \frac{n_\sigma \hat{R} \hat{T} \hat{\Gamma}_m}{\hat{\mu} \hat{U}} \frac{\hat{U}\hat{h}}{\hat{D}_I} = 
   \frac{ \hat{\Gamma}_s n_\sigma \hat{R} \hat{T} \hat{h} }{\hat{\mu} \hat{D}_I}.
\end{equation}
Therefore, in the insoluble case, the plastron slip length is simply inversely proportional to $Ma_\text{ins}$, such that $Ma_\text{ins} \rightarrow \infty$ yields zero slip ($\lambda_{x_0}\rightarrow 0$), whereas $Ma_\text{ins} \rightarrow 0$ allows the plastron to be free-slip ($\lambda_{x_0}\rightarrow \infty$), analogously to the soluble case.}

\paolo{In general, computing the effective slip length (or equivalently the drag) requires going thorugh the Stokes flow calculation described in \S\ref{sec:transverseRidges}. However, in the $g\gg 1$ limit, it is also possible to evaluate analytically the effective slip length $\lambda_e$, and therefore the drag reduction. We start from~(\ref{eq:globaleffsliplength}), which expresses $\lambda_e$ as a function of $\gamma_{Ma}$ and $E_0$. Note that, based on (\ref{eq:C_for_k_0}),
\begin{equation}
\left.E_0\right|_{g \gg 1} =  \frac{ \left. E^{(0)} \right|_{g \gg  1} }{ (1-\gamma_{Ma})} \simeq \frac{\phi}{(4-3\phi)},
\end{equation}
yielding $\lambda_e$ in terms of $(\gamma_{Ma},\phi)$
\begin{equation}\label{eq:lambda_e_gLarge}
   \left. \lambda_e \right|_{g \gg 1} \simeq \frac{2\phi(1-\gamma_{Ma})}{(4-3\phi) - \phi (1-\gamma_{Ma})}.
\end{equation}
To calculate $\gamma_{Ma}$ in the insoluble limit, use~(\ref{eq:uI_ins}) to eliminate $u_I$ in~(\ref{eq:uIscaled1}), and obtain
\begin{equation}\label{eq:gamma_ins_gLarge}
   \left. \gamma_{Ma} \right|_\text{ins} \simeq \frac{ Ma_\text{ins} \mathcal{F}_0}{1+ Ma_\text{ins} \mathcal{F}_0}.
\end{equation}
For $\mathcal{F}_0 = u_{Ic}/[2(1-\gamma_{Ma})]$, use the large-$g$ approximation~(\ref{eq:uicScaling3}), that is $\mathcal{F}_0\simeq 1/(4-3\phi)$. Substituting into (\ref{eq:gamma_ins_gLarge}) and then into (\ref{eq:lambda_e_ins_gLarge}), the effective slip length for insoluble surfactant over a long grating is found explicitly as
\begin{equation}\label{eq:lambda_e_ins_gLarge}
   \left. \lambda_e \right|_{\text{ins},g \gg 1} \simeq \frac{2\phi}{ Ma_\text{ins}+4(1-\phi)}.
\end{equation}
}

\paolo{For long gratings, analytical expressions for $\lambda_{x_0}$ and $\lambda_e$ are also possible in the case of soluble surfactant. If $g\gg 1$, we expect the diffusive boundary layer to be limited by the channel height, and therefore $\delta$ will approach a constant. From our simulations, we find $\delta \simeq 1.20$ in this limit. For the plastron slip length, if $g$ is sufficiently large, we expect the second term in~(\ref{eq:sliplengthmidgap}) to be dominant, yielding
\begin{equation}
   \left. \lambda_{x_0} \right|_{g \gg 1} \simeq \frac{2 a_2}{a_1\, k^*} \frac{g^2\,Bi\,\chi}{\chi+1.20\,Bi\,Pe}.
\end{equation}}
\paolo{To find $\lambda_e$, we calculate $\gamma_{Ma}$ using~(\ref{eq:gammaMa}), where we again set $\delta \simeq 1.20$ and $\mathcal{F}_0\simeq 1/(4-3\phi)$. Without further approximation we obtain
\begin{equation}\label{eq:gamma_gLarge}
   \left. \gamma_{Ma} \right|_{g \gg 1} \simeq \frac{a_1 k^* Pe_I (\chi + Bi\,Pe)}{(\chi + Bi\,Pe)\left[(4-3\phi)+a_1k^*Pe_I\right] + 1.20\,a_2(4-3\phi)g^2 Bi\,\chi Pe_I}.
\end{equation}}
\paolo{Recalling that $a_1 = 2.30$ and $a_2 = 0.319$, equations (\ref{eq:gamma_gLarge}) and (\ref{eq:lambda_e_gLarge}) together provide explicitly the effective slip length as a function of surfactant properties and geometry, without the need to solve the full Stokes flow problem. The drag reduction is then found from (\ref{eq:DR}), as before.
}

\vspace{5pt}
\subsection{\paolo{Tentative deductions for turbulent regimes}}
Applications of our model in turbulent regimes might be possible if a sufficiently thick viscous sublayer exists. If the surfactant transport occurs within the viscous sublayer, \julien{where the flow is laminar}, the viscous sublayer height would be the appropriate length scale instead of $\hat{h}$, \paolo{and the flow velocity at the edge of the viscous sublayer would be the relevant velocity scale $\hat{U}$.} In that case, the local characteristic velocity $\mathcal{F}_0 \hat{U}$ at the interface may be sufficiently small to avoid the partial SC regime. \paolo{While the resulting predictions \julien{based on our model} would be at best qualitative, it is of great practical interest to explore this tentative application to turbulent flows. Here we restrict ourselves to examining the plastron slip length $\hat{\lambda}_{x_0}$, \julien{defined in \eqref{eq:sliplengthmidgap}  and which does not depend on whether the flow is internal (\eg channel flow) or external (\eg a boundary layer)}.}

\paolo{In a turbulent boundary layer, with dimensional wall shear stress $\hat{\tau}_w$, the canonical scales are the shear velocity $\hat{u}_\nu = \sqrt{\hat{\tau}_w/\hat{\rho}}$ and the viscous length scale $\hat{\delta}_\nu = \hat{\nu}/\hat{u}_\nu$~\citep{Pope2000-qq}. The height of the viscous sublayer is of order $10\,\hat{\delta}_\nu$. At this distance from a smooth wall, the flow velocity is of order $10\,\hat{u}_\nu$~\citep{Pope2000-qq}. We replace $\hat{h}$ and $\hat{U}$ in our analysis with these turbulent scales and set a representative wall shear stress $\hat{\tau}_w = 50\,$N/m$^2$.}

\begin{figure}
\centering
\includegraphics[width=0.95\linewidth]{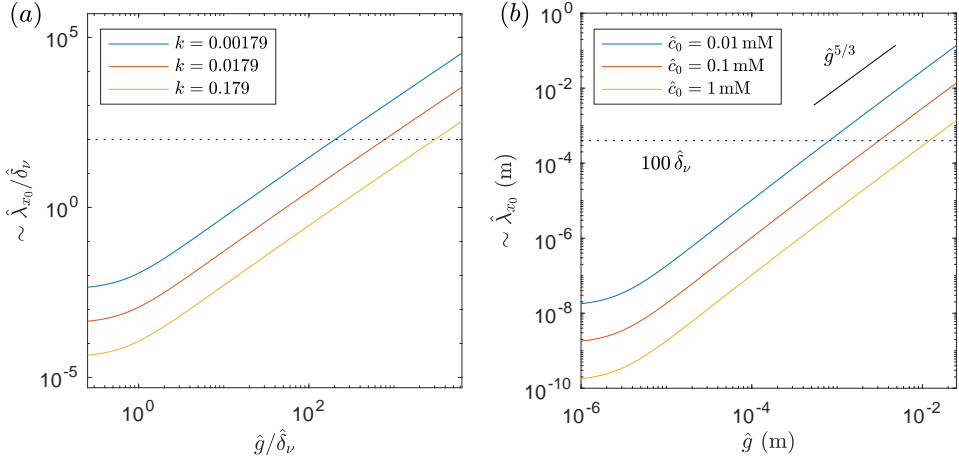}
\caption{\paolo{Order-of magnitude scaling for the slip length over the plastron of turbulent superhydrophobic gratings, such as those considered by~\cite{Park2014-an}, as a function of the grating length $\hat{g}$. A prescribed shear stress of $50\,$N/m$^2$ is used in the calculation. ($a$) shows the mid-plastron slip length normalized by the viscous  lengthscale $\hat{\delta}_\nu$, whereas ($b$) shows the corresponding dimensional results.}
}
\label{fig:turb}
\end{figure}

\paolo{In practical applications, detection of a specific surfactant type is challenging. However, dimensional surface tension $\hat{\sigma}$ has been measured for both clean `synthetic' seawater (labelled `$\hat{\sigma}_0$' below), as well as for seawater samples collected through cruises~\citep[][and references therein]{Nayar2014-yo}. For the purpose of estimating the order of magnitude of $k$ in this example, we use the Langmuir isotherm. While this is less accurate than the Frumkin isotherm, it does not require $k\ll1$, yet it provides a relation between surfactant and surface tension that can be analytically inverted. In a liquid at equilibrium~\citep{Chang_Franses_1995}
\begin{equation}
\hat{\sigma}_0-\hat{\sigma} = n_\sigma \hat{R} \hat{T}\hat{\Gamma}_m \ln{\left(1+k\right)}. \label{eq:sigmaLangmuir}    
\end{equation}
\cite{Schmidt2011-tu} find that seawater that is away from major surfactant sources (such as seasonal blooms of phytoplankton, oil seeps or wastewater treatment facilities) has $\hat{\sigma}_0-\hat{\sigma} \sim 10^{-4}$\,N/m \cite[see also][]{Pogorzelski2001-ie}. Setting $n_\sigma \approx 2$, $\hat{R}=8.314$\,kg\,m$^2/$(s$^2$\,K\,mol), $\hat{T}\approx 300$\,K,  $\hat{\Gamma}_m \approx 3.9\times 10^{-6}$\,mol$/$m$^2$ and rearranging (\ref{eq:sigmaLangmuir}) for $k$, we obtain, for low-surfactant oceanic conditions
\begin{equation}\label{eq:kInOcean}
k =  \exp\left( \frac{\hat{\sigma}_0-\hat{\sigma}}{n_\sigma \hat{R} \hat{T}\hat{\Gamma}_m}  \right) - 1 \approx 0.005.
\end{equation}
Incidentally, this example yields $k \ll 1$, consistently with our set of assumptions. Note that substantially higher $k$ values can occur in oceans and lakes.} 
\paolo{In order to set up a well-defined calculation, we consider SDS with concentrations $\hat{c}_0 = (0.01, 0.1, 1)$\,mM, corresponding to $k=1.79\times (10^{-3}, 10^{-2}, 10^{-1} )$, which bracket the value of $k$ found in~(\ref{eq:kInOcean}).} 
\paolo{We change the length of the grating $\hat{g}$ from $\SI{1}{\micro \meter}$ to $\SI{2.5}{cm}$, the latter being the grating length in~\cite{Park2014-an}. We use~(\ref{eq:sliplengthmidgap}) to calculate $\hat{\lambda}_{x_0}$, as shown in figure~\ref{fig:turb}.}

\paolo{To interpret figure~\ref{fig:turb}, we note that one needs the effective slip length to be comparable to the thickness of the viscous sublayer in order to achieve meaningful drag reduction in turbulent flow~\citep{Rothstein:2010im}. For this substantial effective slip to be possible, one needs the plastron to have an even larger slip length, since of course the solid walls will have no-slip. (For context, recall that, in canonical \julien{surfactant-free} theories and simulations, the plastron is assumed to have infinite slip length.) Since the viscous sublayer thickness is $10\,\hat{\delta_\nu}$, we propose that a plastron slip length of around $100\,\hat{\delta_\nu}$ is a tentative relevant threshold for useful drag reduction. This value is marked by a dashed line in \fernando{figure \ref{fig:turb}}.
}

\paolo{Note that, at small grating lengths $\hat{g}$, the first term in the right-hand side of ($\ref{eq:sliplengthmidgap}$) dominates. This is independent of $\hat{g}$. At larger $\hat{g}$, the second term in ($\ref{eq:sliplengthmidgap}$) dominates, eventually following a scaling of  $\hat{g}^{5/3}$, as shown in figure~\ref{fig:turb}($b$). The slip length also increases with the inverse of $\hat{c}_0$. These results suggest that useful drag reduction may be possible provided the surfactant concentration is not too strong and the plastron is sufficiently long in the streamwise direction. Our conclusions are consistent with the experimental results of~\cite{Park2014-an}, who found strong drag reduction for gratings in laboratory experiments, indicating that traces of surfactant were not sufficient to negate drag reduction. However, our theory also indicates that a large drag increase may occur for a ship equipped with SHS, when it navigates through surfactant-rich waters, which are common in the coastal ocean, rivers, and lakes. Finally, we emphasize, once again, that these are tentative deductions, and that our model will require additional work to provide quantitative drag predictions in turbulent flow.}

\subsection{\paolo{Relative importance of effects neglected in the present model}}

\subsubsection{\paolo{Surface rheology}}
\fernando{It is also worth  discussing some physical effects not considered in our model. For example, surface rheology could play a role in the boundary condition \eqref{eq:MarangoniStress} if viscous surface stresses at the interface were comparable to viscous stresses  in the bulk. The relevant dimensionless groups accounting for this balance are the Boussinesq numbers $Bo_\mu = \hat{\mu}_s/\hat{\mu}\hat{g}$ and $Bo_\kappa = \hat{\kappa}_s/\hat{\mu}\hat{g}$, with $\hat{\mu}_s$ and $\hat{\kappa}_s$ the surface shear and the surface dilatational viscosities of the surfactant-laden interface, respectively. The precise measurement of $\hat{\mu}_s$ and $\hat{\kappa}_s$ is itself a challenging problem with many open questions \citep{Langevin14annrev}.}

\fernando{A recent experimental study by \cite{Zell14}, who employed a technique of unprecedented precision, concludes that \emph{soluble} surfactants can be regarded as surface shear inviscid, with values of $\hat{\mu}_s$ below their experimental sensitivity of $\SI{e-8}{\kg \per \s}$. In our problem, 
we can expect a negligible effect from surface shear viscous stresses, even at the smallest practical SHS length $\hat{g}$. Indeed, assuming a worst-case scenario with $\hat{\mu}_s=\SI{e-8}{\kg \per \s}$ in water ($\hat{\mu}\approx\SI{e-3}{\kg \per \m \per \s}$) we find $Bo_{\mu}\ll{1}$ for $\hat{g}\gg{\hat{\mu}_s/\hat{\mu}}=\SI{e-5}{\m}$, which is the case in practical applications.} 
%

\fernando{Surface dilatational viscosities are even more challenging to measure, since dilatational rheology and Marangoni stresses are necessarily coupled, and therefore hard to distinguish, at an interface subject to compression or expansion (\cite{elfring16,Kotula15}). However, a natural (although unverified) assumption for soluble surfactant is to assume  $\hat{\kappa}_s\sim\hat{\mu}_s$ \citep{Langevin14}, leading to $Bo_\kappa\sim{Bo}_\mu$. Thus,  surface dilatational viscous stresses can also be considered negligible for SHS geometries with practical gap lengths $\hat{g}\gg\SI{e-5}{\m}$. 
Note also that it is common to find an \emph{effective} surface dilatational viscosity in the literature which can be much larger than $\hat{\mu}_s$. However, unlike the true \emph{intrinsic} viscosity $\hat{\kappa}_s$, the effective surface dilatational viscosity actually accounts for dissipation from other non-rheological effects such as adsorption--desorption fluxes, which are already accounted for explicitly in our study.}

\subsubsection{\paolo{Viscous stresses in the gas phase}}
\julien{In this model, we have  neglected   viscous stresses from a gas phase inside the grating compared with the other stresses, namely the driving viscous stress from the liquid phase and the surfactant-induced Marangoni stress  at the interface. To assess the validity of this assumption, we compare an order-of-magnitude estimate of the characteristic  gas viscous stress with order-of-magnitude estimates of the other two stresses.} 

\julien{Let us consider the condition of  continuity of stress at a surfactant-free interface of a SHS, where a viscous gas phase fills two-dimensional rectangular gratings of depth $\hat{H}_g$ \cite[see \eg][]{schonecker13,crowdy17apr}. The gas viscous stress at the interface, normalised by the characteristic driving stress from the liquid phase, is at most of the order of $\epsilon u_{Ic}/H_g$, where $u_{Ic}=\hat{u}_{Ic}/\hat{U}=2\mathcal{F}_0(g,\phi)$ is the maximum shear-free interfacial velocity computed using \eqref{eq:uicScaling0} at $x=0$ and with $\gamma_{Ma}=0$, $\epsilon=\hat{\mu}_{g}/\hat{\mu}$ is the dynamic viscosity ratio between the gas and liquid phases, and $H_g=\hat{H}_g/\hat{h}$ is the normalised depth of the grating. From all our simulations presented in figures~\ref{fig:lambdaEffTheoryVSData}--\ref{fig:gammaMaTheoryVSData}, we estimate that the gas viscous stress is negligible compared with the  driving stress from the liquid phase, $\epsilon u_{Ic}/H_g\ll 1$, for all $\hat{H}_g\gtrsim \SI{e-5}{ m}$, except at high viscosity ratio $\epsilon \gtrsim 1$. To calculate $\epsilon$ we have assumed that the gas in the grating is air, $\hat{\mu}_g=\SI{1.81 e-5}{\kg \m^{-1} \s^{-1}}$, whilst the liquid viscosity varies over a broad range such that $\num{1.81e-4}\leq \epsilon \leq \num{1.81e5} $. 
}

\julien{Compared with the  Marangoni stress measured in our simulations, which can be estimated as $k^*/g$ when normalised with the driving viscous stress from the liquid phase (see \eqref{eq:MarangoniStresslin} and \eqref{eq:concexpansion}), the gas viscous stress is also negligible in all our simulations, $(\epsilon u_{Ic}/H_g)/(k^*/g)\ll 1$ for all $\hat{H}_g\gtrsim \SI{e-5}{m}$ and all $\epsilon$. If we assume $\hat{H}_g\sim \SI{e-6}{ m}$, we find that the gas viscous stress is of the same order of magnitude as the Marangoni stress in a small number of  simulations only.
}

\julien{We have also studied the effect of air viscosity in the experiments of \cite{peaudecerf17} and \cite{Song2018-uw}, who measured the velocity profile near the air--water interface of SHSs made of longitudinal rectangular gratings in laminar channel flows. We find that the normalised air viscous stress at the interface is approximately $\epsilon u_{Ic}/H_g\approx 0.05$ in the experiments of \cite{peaudecerf17} and of the order of $0.001$ to 0.01 in the experiments of \cite{Song2018-uw}. This ratio falls by at least an order of magnitude when using their measured (reduced) slip velocity $u_I$, instead of our theoretical shear-free prediction \eqref{eq:uicScaling0}. Since air viscous stresses are typically several orders of magnitude smaller than the characteristic driving force from the water phase, air viscous effects alone cannot explain the negligible or reduced slip velocity measured in their experiments. As we have shown above, the presence of surfactants, as modelled in this study, provides a consistent explanation for the reduced or negligible slip they measured.}

\julien{The experimental and numerical study of \cite{Schaffel2016-mh} also provides compelling evidence that viscous effects from a gas phase are generally negligible or second order effects. They  report experimental local and effective slip lengths  in  microfluidic channel flows over  a SHS made of pillars. Since their geometry  differs from the rectangular gratings considered in our model, the effect of Marangoni stresses due to the presence of surfactant is more difficult to estimate. Any surfactants in their experiments are transported over a complex two-dimensional interface with multiple local stagnation points, rather than a one-dimensional interface with two clear stagnation points. Nevertheless, their comparison  between  experimentally measured slip lengths and the slip lengths obtained from numerical simulations
is revealing. Using their notation, the local experimental slip lengths, named $b_{local,exp}$ (see their \fig~2b), is  approximately 7\% to 93\% lower (depending on the location at the interface) than the slip lengths obtained numerically, which already account for viscous effect from the gas phase (named $b_{cp}$, see their \fig~4b). Moreover, they find that if viscous effects from the gas phase are neglected in the numerical simulations, the effective (global) slip length increases only slightly, from $b_{eff,th}=\SI{4.0}{\micro\meter}$ to $\SI{4.3}{\micro\meter}$, compared with  $b_{eff}=\SI{1.7}{\micro\meter}$ as measured experimentally. They attribute the 58\% reduction in the experimental effective slip length to `interface contamination', \ie surfactant, explaining that viscous effects from the gas phase cannot explain the discrepancy with their numerical simulations.
}

\julien{Based on our own simulations and the studies of \cite{Schaffel2016-mh}, \cite{peaudecerf17}, and \cite{Song2018-uw}, we find that viscous effects from a gas phase inside SHS gratings can generally be  neglected for most practical applications, as intuitively expected and commonly assumed in the SHS literature. Indeed, in many applications or experimental studies on SHS, the liquid and gas phases are often water and air, respectively, such that $\epsilon\approx 0.02$ is very small. Moreover, the grating depth $\hat{H}_g$ can often  be made sufficiently large so as to minimize viscous effects from the gas phase. The criterion found based on our simulations is $\hat{H}_g\gtrsim \SI{e-5}{m}$, which is technically feasible in many applications \fernando{and often necessary in experiments to prevent collapse of the plastron during the filling of the chamber}. We note that in general this criterion depends  on the geometry, such as the ratios $H_g$ and $H_g/g$, and whether the flow is confined in a channel or unbounded in a semi-infinite domain. 
For further detail about viscous effects from the gas phase in surfactant-free SHS flow, we refer the reader to the   theoretical and numerical studies of \cite{schonecker13}, \cite{schonecker14} and \cite{crowdy17apr}. These studies also show that  gas viscous effects are mainly important at large $\epsilon$ or  small $H_g$,  consistently with our findings. In these particular regimes, both viscous effects from the gas phase and surfactant Marangoni stresses would need to be modelled in order to assess their respective contribution on the drag reduction of the SHS.
}

\subsubsection{\paolo{Interface deformation}}\label{sec:discInterfaceDeformation}
\julien{Another physical mechanism not considered in the present study is the effect of interface deformation. Many studies have investigated the effect of lateral or longitudinal curvature of the air--water interface of SHSs \cite[see \eg][and references therein]{crowdy17apr,kirk17,game19}. They have found  positive or negative impact depending on the curvature sign (whether it points towards the liquid phase or the gas phase), geometry (transverse or longitudinal SHSs),  whether the flow is bounded or unbounded \citep{kirk17}, or the Reynolds number \citep{game19}. 
}

\julien{The deformation of the interface could be due to the gravity force, viscous forces or a  pressure difference across the interface. These forces must be compared with the surface tension\fernando{, which resists deformations associated with an increase in surface area of the interface, \ie flattening the plastron in this specific problem}. In general, gravity can be neglected since the  smallest length scale in microfluidic applications is much smaller than the capillary length\fernando{, which in this case has a typical value of $\hat{l}_c =\nobreak \sqrt{\hat{\sigma}_0/(\Delta\hat{\rho}\,\hat{a}_g)} \approx \SI{2.7e-3}{\m}$. For this estimate we have chosen the representative values of $\hat{\sigma}_0\approx\nobreak \SI{7.2e-2}{\N \per \m}$ for the surface tension, $\hat{a}_g=\SI{9.81}{\m \per \s^2}$ for the gravitational acceleration, and $\Delta\hat{\rho} \approx \SI{1000}{\kg \per \m^3}$ for the air-water density difference.} Similarly, viscous forces are neglected in most applications due to small  capillary numbers near the interface $Ca_I= u_I Ca=\hat{\mu} \hat{u}_I/\hat{\sigma}$. We typically find $Ca_I\lesssim \SI{e-3}{}$ for a surfactant-free air--water interface across our range of parameters. We note the effect of surfactant would on the one hand tend to reduce $\sigma$, thus enhancing interfacial deformation. On the other, as we have shown in this study, surfactant would  reduce $u_I$, thus \francois{limiting interfacial deformation}. The reduction of $u_I$ can occur at concentrations much smaller than concentrations necessary to change the surface tension noticeably \cite[][]{Chang_Franses_1995}. Hence, we intuitively expect that viscous forces would have negligible effect, even when combined with surfactant, in regimes where surfactants affect $u_I$. The capillary length and the capillary number depend on the properties of the fluids on either side of the interface. Although air--water systems, as assumed above, are the most common across real applications,  laboratory and field experiments, these characteristic numbers would need to be examined carefully in more specialised applications \cite[\eg liquid metals for micro-cooling,][]{lam15}.
}

\julien{The effect of  pressure difference is one of the most common cause of interfacial deformation \cite[\eg][]{game19}. Interfacial curvature typically depends on the ratio of the pressure difference and the surface tension, $\Delta p/\sigma$, following the Young--Laplace law. Thus, surfactant could enhance curvature by reducing $\sigma$, thereby affecting the performance of the SHS. Similar to what we noted for viscous effect, we expect the negative impact of surfactant on $u_I$ via Marangoni effects to be generally more important than via interface deformation. 
If the  pressure difference is large enough, there can exist some regimes where both interface deformation and Marangoni stresses are important. The combined effects on SHS performance of negative Marangoni effects and positive or negative interfacial deformation effects would be an important topic for future research.
}

\subsubsection{Three-dimensional effects}
Although the geometry used in our model is two-dimensional, we expect the model to give a reasonable estimate of the impact of surfactants for flows above three-dimensional rectangular longitudinal SHS gratings, similar to those used by \cite{peaudecerf17} and many other studies. %
For three-dimensional gratings with small aspect ratio $w/g = 1/15$, \cite{Song2018-uw} observed three-dimensional flows with recirculations along the side boundaries or via the interior, depending on whether the interface was convex or concave. Overall, they found significant reduction of the slip velocity at the interface due to surfactant contamination, which shows that these three-dimensional recirculation flows are secondary effects compared to the mean two-dimensional effects due to the surfactant-induced Marangoni forces.
For cases without this recirculation pattern, we expect surfactants to be advected along the grating, forming a longitudinal surfactant gradient which is approximately uniform in the spanwise direction (\ie across the grating width). Owing to spanwise viscous friction, we can also note that our model would give a lower bound prediction on the surfactant-induced Marangoni shear, or conversely, an upper bound for the effective slip length and maximum drag reduction.

\section{Conclusions}\label{sec:conc}
In this study, we present a reduced-order scaling model to account for the impact of soluble surfactants in channel flows with superhydrophobic surfaces. The drag reduction potential of superhydrophobic surfaces can be severely reduced if surfactants adsorbed onto the plastron 
induce Marangoni forces opposed to the flow. \paolo{These Marangoni forces develop when a gradient of surfactant establishes along the interface.}

To simplify the governing equations of this problem, we first linearised the kinetics source terms for the surfactant flux between the bulk and the interface, as well as the coupling condition balancing the viscous force and the surfactant-induced Marangoni force. This linearisation holds for small surfactant concentration $\Gamma\ll 1$, which is a reasonable assumption for most applications where surfactants are not artificially added.
Then, integrating the transport equations in the bulk and at the interface, we find a linear relationship between the interfacial slip velocity at mid-gap and the interface-averaged surfactant-induced Marangoni shear, given by (\ref{eq:sliplengthmidgap}). This relationship depends explicitly on the non-dimensional numbers $k^*=k Ma$, which combines both the non-dimensional bulk background surfactant concentration $k$ and the Marangoni number $Ma$, as well as  $Pe$, $Pe_I$, $g$, $Bi$ and $\chi$.

%
To obtain a global effective slip length and predict how surfactant transport can affect the flow rate and the drag reduction potential of the SHS, we solve the continuity and momentum conservation equations for  low Reynolds number flow. Using a technique based on the work of \cite{Lauga_Stone_JFM_2003} for surfactant-free SHS flow, we solve  Stokes' equation with mixed boundary conditions and a prescribed shear profile at the interface. 
In the case of a uniform interfacial shear  $\gamma_{Ma}$, the interfacial velocity relates linearly to $1-\gamma_{Ma}$, where the coefficient of proportionality  depends on the geometric non-dimensional parameters of the SHS, namely the grating length $g$ and the gas fraction $\gf$. 
We close the problem and eliminate the interface velocity by using our earlier result, based on the surfactant problem, that also related interface velocity to shear.
Hence, we find that the average Marangoni shear $\gamma_{Ma}$ depends on  seven non-dimensional parameters: $k^*$, $Pe$, $Pe_I$, $Bi$, $\chi$, $g$ and $\phi$, following (\ref{eq:gammaMa}). The dependence on the geometry is implicit through the function $\mathcal{F}_0(g,\gf)$, which can be solved from the linear problem (\ref{eq:linsysmatrix}) assuming a surfactant-free Stokes' flow in the same geometry. 
We find that the effective slip length is $\lambda_e = 2(1-\gamma_{Ma})E_0/(1-(1-\gamma_{Ma})E_0)$, see (\ref{eq:globaleffsliplength}), where $E_0=Q_{d,0}/2$ with $Q_{d,0}$ the added volume flow rate in an SHS channel flow without any surfactant. 
The corresponding  added flow rate $Q_d$ and drag reduction $DR$ due to the SHS, in the general case of a surfactant-contaminated flow, can be determined from the effective slip length following (\ref{eq:lambdaeQd}) and (\ref{eq:DR}), respectively. These equations show how the slip length, the added flow rate and the drag reduction are affected by the surfactant-induced Marangoni shear rate at the interface.

In order to test the regime of validity and the accuracy of our model, we performed 137 finite-element numerical simulations of the full governing equations in steady, pressure-driven, laminar channel flows, inclusive of soluble surfactants following (\ref{eq:divu})--(\ref{eq:SurfNoFlux}). We varied the governing non-dimensional groups across a broad range of values to explore the vast parameter space of this problem (see \fig~\ref{fig:parameterOverview}, appendix~\ref{apdx:keydimensionlessnumbers}, table~\ref{tab:NDNminmax} and the Supplementary Table S1). The model predictions for $\lambda_e$, $DR$ and $\gamma_{Ma}$ follow  well the numerical results across almost all the parameter space  explored. The model coefficients are determined through a least-squares fit for $\lambda_e$, yielding $a_1={2.30}$, $a_2 = {0.319}$, $\delta_{0,3}={1.68}$ and $\delta_{1,3}={0.0528}$. 
The flows that are least well captured by our model corresponds to the `partial stagnant cap regime', which is also found in air bubbles rising in surfactant-contaminated water. This regime occurs at very large $\mathcal{F}_0 Pe_{I,g}$, and low $\mathcal{D}_{I,g}$ or low $\mathcal{K}_{I,g}$. The partial SC regime exhibits a sharp increase in the shear rate at the transition between a shear-free upstream part and a no-slip downstream part of the interface, which differs from our assumption of a uniform Marangoni shear along the interface.
Nevertheless, at least for the simulations performed here, our model predictions are sufficiently accurate for practical purposes. It will be important to test the accuracy of our model also in more complex flows.

Canonical SHS models, which completely neglect surfactant effects, can yield a large error in the prediction of the slip length and of the drag reduction, as shown in \fig{}s~\ref{fig:lambdaEffTheoryVSData} and \ref{fig:DRTheoryVSData}. In particular, the error is very large, by several orders of magnitude, at large Marangoni stresses. Hence, models neglecting surfactant can significantly overestimate the drag reduction potential of the SHS. This is particularly important in applications where small background environmental surfactant traces are sufficient to induce strong Marangoni forces, as previously found by \cite{peaudecerf17}.

Overall, the model we present provides a useful quantitative estimate of the effect of surfactants on the drag reduction potential of SHSs, across a vast part of the parameter space except in the partial stagnant cap regime. Our scaling predictions can be used directly  in  numerical simulations of flow over SHS in realistic conditions where surfactants cannot be neglected. The effective slip length $\lambda_e$ can be used in a Navier-slip boundary condition on the SHS side, without having to solve the full coupled nonlinear surfactant transport problem.  This will reduce considerably the computational burden associated with realistic simulations of SHS flows.
We also note that our model can be easily adapted for a two-sided SHS channel, via changes in the boundary conditions in the Stokes' flow problem (see \S\ref{sec:StokesFlowModel}). This change in boundary conditions will modify the geometric function $\mathcal{F}_0$. 

Future work will investigate how the model can be modified for more complex three-dimensional flows over SHSs, such as pillars or disordered SHSs. Apart from annular flows \citep{Lee2008-mg,Song2018-uw} or very long air--water interfaces \citep{peaudecerf17}, accumulation of surfactant at stagnation points in these three-dimensional problems can also lead to surfactant-induced Marangoni stresses. Predicting the magnitude of these forces  and the overall effect on the effective slip length or the drag reduction is a complex problem.
Many applications operate at larger Reynolds numbers, where the effect of turbulence on the surfactant Marangoni stresses may be important. At intermediate Reynolds numbers, where the viscous sub-layer forming at the SHS is sufficiently thick compared with the surfactant diffusive boundary layer thickness, our scaling model may still be applicable, though the empirical parameters may differ from those found here. At very large Reynolds numbers, turbulence is likely to enhance the diffusion of surfactant in the bulk and at the interface, which could  change  the concentration gradients and result in intermittent localised Marangoni forces at the interface. These problems have a direct impact on the performance of SHSs in many applications, and constitute important topics for future studies.


\section{Acknowledgments}

We gratefully acknowledge financial support from the Raymond and Beverly Sackler Foundation, the Engineering and Physical Sciences Research Council, the European Research Council Grant 247333, Mines ParisTech, the Schlumberger Chair Fund, the California NanoSystems Institute through a Challenge Grant, ARO MURI W911NF-17-1-0306 and ONR MURI N00014-17-1-2676.

\nolinenumbers


\fancyhead[r]{}
\fancyhead[l]{}
\bibliographystyle{unsrt}
\bibliography{Biblio-LaminarSHSmodel}



\clearpage
\appendix





\fancyhead[r]{}
\fancyhead[l]{}

\section{Key dimensionless numbers across all numerical simulations}\label{apdx:keydimensionlessnumbers}
\begin{figure}
\centering
\includegraphics[width=0.8\linewidth]{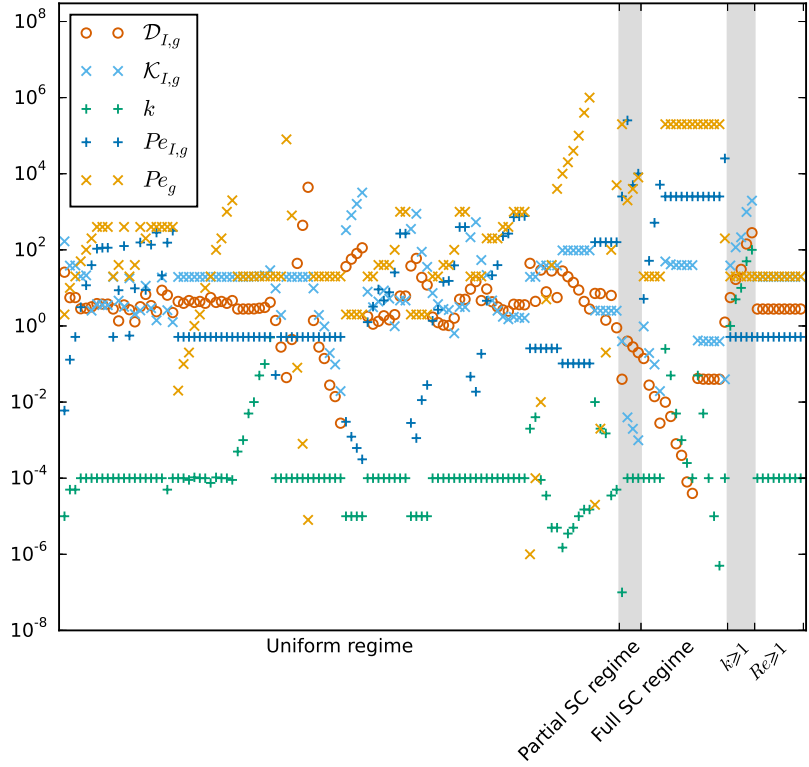}
\caption{
Variation of some of the characteristic non-dimensional numbers used in our 137 numerical simulations depending on the regime. The non-dimensional number $\mathcal{K}_{I,g}=Bi_g(1+k)/\mathcal{F}_0$ the ratio of the adsorption--desorption kinetics flux to the advective flux at the interface. The parameter $\mathcal{D}_{I,g}=\chi_g(1+k)/(\mathcal{F}_0 Pe_g)^{1/2}$ is the ratio of the transverse diffusive flux through the diffusive boundary layer to the advective flux at the interface. The function $\mathcal{F}_0=\mathcal{F}(g,\gf,x=0)$ is related to the interfacial slip velocity following (\ref{eq:uIscaled1}).
}
\label{fig:parameterOverview}
\end{figure}

%
%
To help provide a visual overview of the simulations performed, \fig~\ref{fig:parameterOverview} plots the value of each dimensionless group on the vertical axis, with the horizontal axis indicating different simulations. Ranges for each parameters are also reported earlier in table~\ref{tab:NDNminmax}. Detailed values are included in table~S1 of the Supplementary Materials.

\section{Diffusive boundary layer thickness}\label{apdx:DiffBLthickness}

To determine an estimate of the  boundary layer thickness $\delta$ for the surfactant concentration, we builds on the result in \S\ref{sec:scalingTheory} and perform a scale analysis of the bulk advection--diffusion equation (\ref{eq:Cstar}), which is expanded below as
\begin{equation}\label{eq:Cstarapdx}
u\fp{c}{x}{} + v\fp{c}{y}{} =  \frac{1}{Pe} \left(\fp{c}{x}{2} + \fp{c}{y}{2} \right).
\end{equation}
In the surfactant  adsorption (resp. desorption) boundary layer forming above the interface, we denote the characteristic variation of the bulk concentration as $\Delta c$. In the streamwise 
direction, we expect the change $\Delta c$ to take place between between $x=-g/2$ and $x=x_0$ (resp. $x_0$ and $g/2$), as sketched in figure~\ref{fig:schematicSHSchannel}. As explained in \S\ref{sec:scalingTheory}, $x_0$ is defined as the interface location where the kinetics flux $S$ vanishes. Under the assumption of low interfacial concentration 
(see (\ref{eq:concexpansion}) and text above), we previously found that the adsorption and desorption diffusive boundary layers are approximately anti-symmetric and of characteristic streamwise length scale $\sim g$, as depicted in figure~\ref{fig:schematicSHSchannel}(\textit{b}). 

If we focus on the adsorption region of the interface, $c$ at the interface is denoted as $c_I$, which varies by a scale $\Delta c_I$ between $x=-g/2$ and $x=x_0$, where $S=0$ implies $c_I\sim 1$. In addition, the characteristic cross-stream 
variation, across the boundary layer, 
is from $c_I$ to  $1$, implying that this variation in $c$ also scales as $\Delta c_I$. Therefore, in both the $x-$ and $y-$ directions, $\Delta c=\Delta c_I$ over characteristic distances $g$ and $\delta$, respectively.

We denote the  characteristic streamwise and cross-stream velocities in the diffusive boundary layer as $U_{\delta}$ and $V_{\delta}$, respectively. Hence, a scale analysis of  equation (\ref{eq:Cstarapdx}) gives
\begin{equation}\label{eq:ADeqscale}
U_{\delta} \frac{\Delta c_I}{g} + V_{\delta} \frac{\Delta c_I}{\delta} \sim  \frac{1}{Pe} \left( \frac{\Delta c_I}{g^2} + \frac{\Delta c_I}{\delta^2} \right),
\end{equation}
where we can divide throughout by $\Delta c_I$. Thus, $\delta$ is a function of the \Peclet number $Pe$, the interface length $g$, as well as $U_{\delta}$ and $V_{\delta}$. These velocity scales are expected to depend on the geometrical parameters $g$ and $\gf$, as well as on the interfacial velocity $u_I$ and the characteristic shear rate profile $\gamma_{Ma}$. We now seek an explicit dependence of $U_{\delta}$ and $V_{\delta}$ on these parameters. 

We assume that the diffusive boundary layer thickness is not affected by  the channel height, such that $\delta< 1$. We also assume that a diffusive boundary layer above a particular interface is independent from the other interfaces, and thus independent of the gas fraction $\gf$. We retain the dependence on the interface length $g$.
We can distinguish two main limits influencing $U_{\delta}$ and $V_{\delta}$, depending on the boundary condition  at the interface. This can either consist of a finite slip and negligible shear ($u_I > 0$  and $\gamma_{Ma}\ll 1$), or of no-slip and finite shear ($u_I = 0$ and $\gamma_{Ma}\sim 1$). Hence, in general, $U_\delta\sim u_I+\gamma_{Ma}\delta$. 
The two cases are analyzed further below.
\begin{itemize}
    \item First, for $u_I > 0$ and $\gamma_{Ma}\ll 1$, according to the analysis in appendix~\ref{apdx:Asympt_Vel_Plastron}, we have $U_\delta\sim u_I\sim g$ for $g\lesssim 1$, and $U_\delta \sim u_I\sim 1$ for $g\gtrsim 1$. We determine the scale for $V_\delta$ using the continuity equation (\ref{eq:divuStar}), which gives $V_\delta\sim U_\delta \delta/g$. Replacing these velocity scales into (\ref{eq:ADeqscale}), we find
    \begin{equation}\label{eqapdx:deltaPe12}
        \frac{\delta}{g} = \delta_{0,1} \left(1 + \delta_{1,1} g^2 Pe\right)^{-1/2}\quad \textrm{for}\ g \lesssim 1  
    \end{equation}
    and
    \begin{equation}\label{eqapdx:deltagPe12}
         \frac{\delta}{g} = \delta_{0,2} \left(1 + \delta_{1,2} g Pe\right)^{-1/2} \quad \textrm{for}\ g \gtrsim 1,
    \end{equation}
    where $\delta_{0,1}$, $\delta_{1,1}$, $\delta_{0,2}$, $\delta_{1,2}$ are empirical parameters. 
    
    \item Second, for $u_I$ negligible and $\gamma_{Ma}\sim 1$, $U_\delta$ depends only on the ratio of the thickness of the diffusive boundary layer and the channel height: $U_\delta\sim \delta \gamma_{Ma}=\delta$ for $\delta< 1$. This regime is also known as the \Leveque regime \citep{leveque28,landel16}. Note that $V_\delta\sim 0$ in this case. Replacing these velocity scales into (\ref{eq:ADeqscale}), we find the asymptotic behaviour
    \begin{equation}\label{eqapdx:deltagPe13}
        \frac{\delta}{g} = \delta_{0,3} \left(1 + \delta_{1,3} g^2 Pe\right)^{-1/3},
    \end{equation}
    for any $g>0$, and with $\delta_{0,3}$ and $\delta_{1,3}$ two empirical parameters.
\end{itemize}

As noted before, the results (\ref{eqapdx:deltaPe12})--(\ref{eqapdx:deltagPe13}) are valid provided $\delta< 1$, which is satisfied for large enough \Peclet numbers or small enough gap length. For an intermediate regime with partial slip and partial shear, \ie $U_\delta \sim u_I+\gamma_{Ma}\delta$, we expect that the boundary layer thickness has an exponent between $-1/2$ and $-1/3$. The transition between the slip dominated regime, with scaling (\ref{eqapdx:deltaPe12}) or (\ref{eqapdx:deltagPe12}), and the shear dominated regime, with scaling (\ref{eqapdx:deltagPe13}), should be smooth at low Reynolds numbers.


\section{Asymptotic limits for the slip velocity} \label{apdx:Asympt_Vel_Plastron}
The computation of the slip velocity yields distinctive simplified behaviours in the limits of large and small gap length $g$, as evidenced in \fig~\ref{fig:uIc_with_g_and_phi}. In this section, we analytically derive asymptotic limits for the slip velocity profile $u_I(x)$, and confirm their agreement with the numerically computed values at mid-gap from \fig~\ref{fig:uIc_with_g_and_phi}.

We start by considering the so-called \emph{dual series} comprised of equations (\ref{eq:monster1}) and (\ref{eq:monster2})
\begin{subequations}\label{eq:dual_series}
\begin{alignat}{2}
    2E + \sum_{n=1}^{\infty} d_n \alpha_n \cos(k_n x) &= 0 \qquad &&\text{for } g/2 < |x| \leq L/2, \label{eq:dual_seriesa}\\
    E + \sum_{n=1}^{\infty} d_n \beta_n \cos(k_n x) &= 1-\gamma_{Ma} \qquad &&\text{for } |x| < g/2, \label{eq:dual_seriesb}
\end{alignat}
\end{subequations}
\fernando{with $\alpha_n$ and $\beta_n$ defined in equations \eqref{eq:alpha_n} and \eqref{eq:beta_n}, respectively}. From this set of expressions, it is possible to obtain a closed form of the asymptotic behavior of the slip velocity by considering only the leading order of $\alpha_n$ and $\beta_n$ in the relevant limits. This is done in a similar fashion to \cite{Lauga_Stone_JFM_2003} and \cite{teo09}, who derived expressions for the \emph{effective slip length} from the asymptotic behavior of the first coefficient $E$. However, the \emph{slip velocity} depends on the whole set of coefficients, and in this case it is not enough to derive an expression for the first coefficient only. Indeed, recall the form of $u_{I}(x)$ from (\ref{eq:uIallx})
\begin{equation}
    u_{I}(x) = 2E + \sum_{n=1}^{\infty} d_n \alpha_n \cos(k_n{x}).
\label{eq:uIallXAppdx}    
\end{equation}

The derivations of $u_{I}(x)$ and its value at mid-gap $u_{Ic}=u_I(x=0)$ in the limits of large and small $g$ are presented in the next two subsections.

\subsection{Limit of large gap length} \label{sec:subapdx:lim_large_g}

Consider the limit $g\rightarrow\infty$, with the gas fraction $\gf$ fixed. Since $L=g/\gf$, note that this \fernando{case} necessarily implies $L\rightarrow\infty$ as well. \fernando{Note that, due to our choice of the channel height $\hat{h}$ as the length scale for the nondimensionalization, this limit corresponds to a ``narrow'' channel with the top wall close to the plastron.} Consequently, we have
\begin{equation*}
    k_n = \frac{2\pi n}{L} \rightarrow 0,
\end{equation*}
and \fernando{in this limit} $\alpha_n$ and $\beta_n$ \fernando{can be expanded as}
\begin{subequations}\label{eq:alpha_beta_for_k_0}
\begin{alignat}{2}
    \alpha_n &= -\frac{4k_n^3}{3} &&+ O(k_n^5), \label{eq:alpha_beta_for_k_0a}\\
    \beta_n &= -\frac{8k_n^3}{3} &&+ O(k_n^5). \label{eq:alpha_beta_for_k_0b}
\end{alignat}
\end{subequations}
\fernando{Taking into account that $k_n\sim L^{-1}$, the expressions \eqref{eq:dual_series} and \eqref{eq:alpha_beta_for_k_0} lead to the following expansions of the unknown Fourier coefficients} 
\begin{subequations}\label{eq:E_dn_for_k_0}
\begin{alignat}{2}
    E   &= E^{(0)}          && + O(L^{-1}), \label{eq:E_dn_for_k_0a}\\
    d_n &= d_n^{\,(0)}L^{3} && + O(L^2). \label{eq:E_dn_for_k_0b} 
\end{alignat}
\end{subequations}
Substituting \eqref{eq:alpha_beta_for_k_0} \fernando{and \eqref{eq:E_dn_for_k_0}} into \eqref{eq:dual_series} \fernando{we arrive at the leading-order dual series for $E^{(0)}$ and $d_n^{\,(0)}$.} \fernando{After} introducing the changes of variable $\widehat{d}^{\,\,(0)}_n=(2\pi{n})^3\,d^{\,(0)}_n$ and $z=2\pi{x}/L$, \fernando{this dual series yields}
\begin{subequations}\label{eq:dual_series_k_0}
\begin{alignat}{2}
    \frac{3}{2}E^{(0)} - \sum_{n=1}^{\infty} \widehat{d}^{\,\,(0)}_n \cos(nz) &= 0 \qquad &&\text{for } \gf\pi < |z| \leq \pi, \label{eq:dual_series_k_0a}\\
    \frac{3}{8}E^{(0)} - \sum_{n=1}^{\infty} \widehat{d}^{\,\,(0)}_n \cos(nz) &= \frac{3}{8}\left(1-\gamma_{Ma}\right) \qquad &&\text{for } |z| < \gf\pi, \label{eq:dual_series_k_0b}
\end{alignat}
\end{subequations}
\fernando{and its coefficients} can be obtained exactly. Indeed, after integrating (\ref{eq:dual_series_k_0a}) from $\gf\pi$ to $\pi$ and (\ref{eq:dual_series_k_0b}) from $0$ to $\gf\pi$, one can then sum the two expressions and obtain
\begin{equation}
    E^{(0)} = \frac{\gf(1-\gamma_{Ma})}{(4-3\gf)}.
\label{eq:C_for_k_0}    
\end{equation}
The rest of the coefficients $\widehat{d}^{\,\,(0)}_n$ can be retrieved multiplying (\ref{eq:dual_series_k_0}) by harmonics of the form $\cos(mz)$ with $m\in\mathbb{N}$ and $m\geq 1$. Then, applying the same procedure of integration and summation and invoking orthogonality between the functions, we arrive at
\begin{equation}
    \widehat{d}^{\,\,(0)}_n = -\frac{3(1-\gamma_{Ma})}{(4-3\gf)} \frac{\sin(n\pi\gf)}{n\pi}.
\label{eq:d_for_k_0}    
\end{equation}
Using the obtained set of coefficients, one can now evaluate the slip velocity. \fernando{S}ubstituting (\ref{eq:alpha_beta_for_k_0a}) \fernando{and \eqref{eq:E_dn_for_k_0}} in (\ref{eq:uIallXAppdx}), we have
\begin{equation*}
    u_{I}(z) = 2E^{(0)} - \frac{4}{3}\,\sum_{n=1}^{\infty} \widehat{d}^{\,\,(0)}_n \cos(nz)  + O(L^{-1}).
\end{equation*}
Applying (\ref{eq:C_for_k_0}) and (\ref{eq:d_for_k_0}), one subsequently obtains
\begin{equation}
    u_{I}(z) = \frac{2(1-\gamma_{Ma})}{(4-3\gf)}\left[\gf + 2\sum_{n=1}^{\infty}\frac{\sin(n\pi\gf)}{n\pi}\cos(nz)\right] + O(L^{-1}).
\label{eq:uI_for_k_0_with_sum}
\end{equation}
 First, note that for $\gf=0$ the above expression (\ref{eq:uI_for_k_0_with_sum}) yields $u_{I}(x)=0$ \fernando{at leading order,} as one would expect. We then observe that the expression in brackets in (\ref{eq:uI_for_k_0_with_sum}) is the Fourier cosine series of a square wave with value 1 for $|z-2j\pi|<\gf\pi$ and 0 for $\gf\pi<|z-2j\pi|\leq\pi$, where $j\in\mathbb{N}$. Consequently, by virtue of the uniqueness of a Fourier series one has, after undoing the change of variables
\begin{equation}
    u_{I}(x) = 
    \begin{cases}\displaystyle
        \frac{2(1-\gamma_{Ma})}{(4-3\gf)} + O(L^{-1}) & \text{for }  |x| < g/2,\\
        0                                              & \text{for } g/2 < |x| \leq L/2.
    \end{cases}
\label{eq:uI_for_k_0}
\end{equation}
The fact that the slip velocity tends to a constant value \fernando{as} $g\rightarrow\infty$ is expected, due to the confinement effect of the top wall. Indeed, the disparity of horizontal and vertical length scales ($g\gg{1}$) leads to a lubrication regime in which the slip velocity asymptotically tends to a constant along the plastron. \fernando{In such a regime, the velocity field can be approximated as unidirectional in the central ``core region'' following the thin-gap approximation \citep[see for instance the examples in][]{Leal07}.} From (\ref{eq:uI_for_k_0}), the value of the slip velocity at mid-gap $u_{Ic}=u_I(x=0)$ would then yield \fernando{at leading order}
\begin{equation}
    \cfrac{u_{Ic}}{2(1-\gamma_{Ma})} \simeq \cfrac{1}{4-3\gf}\,,
\label{eq:uIc_for_k_0}
\end{equation}
where $u_{Ic}$ has been normalized with $2(1-\gamma_{Ma})$ following Section \ref{sec:transverseRidges}. Notice that this normalization implicitly assumes $0\leq\gamma_{Ma}<1$, however in the case $\gamma_{Ma}=1$ it is  straightforward from (\ref{eq:uI_for_k_0}) that $u_{Ic}=0$ \fernando{at leading order}.

The expression (\ref{eq:uIc_for_k_0}) is plotted in \fig~\ref{fig:uIc_with_g_and_phi}($b$), confirming the trend of the values $u_{Ic}$ computed numerically. Moreover, note that within this asymptotic regime $g\rightarrow\infty$, the expression (\ref{eq:uIc_for_k_0}) leads to the two following limits
\begin{subequations}\label{eq:uIcenter_for_k_0_limits}
\begin{alignat}{2}
    \frac{u_{Ic}}{2(1-\gamma_{Ma})} &\sim 1 \qquad &&\text{for }\gf\rightarrow 1, \label{eq:uIcenter_for_k_0_limits_phi_1}\\
    \frac{u_{Ic}}{2(1-\gamma_{Ma})} &\sim \frac{1}{4}\qquad &&\text{for }\gf\rightarrow 0, \label{eq:uIcenter_for_k_0_limits_phi_0}
\end{alignat}
\end{subequations}
which are corroborated as well by the asymptotic behavior in \fig~\ref{fig:uIc_with_g_and_phi}(a).

\subsection{Limit of small gap length} \label{sec:subapdx:lim_small_g}

Consider now the limit $g\rightarrow{0}$, with the gas fraction $\gf$ fixed. Then, like in the previous case, $g\rightarrow{0}$ necessarily implies $L\rightarrow{0}$ as well. \fernando{This case corresponds to a ``tall'' channel with distant top walls.} We have
\begin{equation*}
    k_n = \frac{2\pi n}{L} \rightarrow \infty,
\end{equation*}
and \fernando{therefore} $\alpha_n$ and $\beta_n$ \fernando{can be expanded as}
\begin{subequations}\label{eq:alpha_beta_for_k_inf}
\begin{alignat}{2}
    \alpha_n &= -e^{k_n}     && + O(e^{-k_n}), \label{eq:alpha_beta_for_k_infa}\\
    \beta_n  &= -2k_ne^{k_n} && + O(e^{-k_n}). \label{eq:alpha_beta_for_k_infb}
\end{alignat}
\end{subequations}
\fernando{Given the functional form of the leading order terms in \eqref{eq:alpha_beta_for_k_inf}, we introduce} the change of variable $\widehat{d}_n=e^{k_n}\,d_n$ \fernando{and seek the expansions}
\begin{subequations}\label{eq:E_dn_for_k_inf}
\begin{alignat}{3}
    E   &= E^{(0)}                       && + E^{(1)}L              && + O(L^2), \label{eq:E_dn_for_k_infa}\\
    \widehat{d}_n &= \widehat{d}_n^{\,\,(0)} && + \widehat{d}_n^{\,\,(1)}L  && + O(L^2). \label{eq:E_dn_for_k_infb} 
\end{alignat}
\end{subequations}
\fernando{After re-scaling the spatial variable $z=2\pi{x}/L$, we insert \eqref{eq:alpha_beta_for_k_inf} and \eqref{eq:E_dn_for_k_inf} into \eqref{eq:dual_series} and group the $O(1)$ terms to arrive at the leading-order dual series for $E^{(0)}$ and $\widehat{d}_n^{\,\,(0)}$}
\begin{subequations}\label{eq:dual_series_leading_k_inf}
\begin{alignat}{2}
    -2E^{(0)} + \sum_{n=1}^{\infty} \widehat{d}_n^{\,\,(0)} \cos(nz) &= 0 \qquad &&\text{for } \gf\,\pi < |z| \leq \pi, \label{eq:dual_series_leading_k_infa}\\
    \sum_{n=1}^{\infty} n\,\widehat{d}_n^{\,\,(0)} \cos(nz)          &= 0 \qquad &&\text{for } |z| < \gf\,\pi,\label{eq:dual_series_leading_k_infb}
\end{alignat}
\end{subequations}
\fernando{which leads to $E^{(0)}=0$ and $\widehat{d}_n^{\,\,(0)}=0$. The terms of order $O(L)$ can then be grouped into the following dual series for $E^{(1)}$ and $\widehat{d}_n^{\,\,(1)}$}
\begin{subequations}\label{eq:dual_series_k_inf}
\begin{alignat}{2}
    -2E^{(1)} + \sum_{n=1}^{\infty} \widehat{d}_n^{\,\,(1)} \cos(nz) &= 0 \qquad &&\text{for } \gf\,\pi < |z| \leq \pi, \label{eq:dual_series_k_infa}\\
                \sum_{n=1}^{\infty} n\,\widehat{d}_n^{\,\,(1)} \cos(nz) &= -\frac{1}{4\pi}\left(1-\gamma_{Ma}\right) \qquad &&\text{for } |z| < \gf\,\pi. \label{eq:dual_series_k_infb}
\end{alignat}
\end{subequations}
\fernando{T}he coefficients \fernando{in \eqref{eq:dual_series_k_inf}} can be obtained exactly \fernando{following the procedure of} \cite{Sneddon66}. \fernando{However, in this case it is not necessary to explicitly obtain $E^{(1)}$ and $\widehat{d}_n^{\,\,(1)}$ in order to obtain the slip velocity, since} the left-hand side of equation (\ref{eq:dual_series_k_infa}) \fernando{can be determined exactly} for $|z| < \gf\pi$ \citep[see page 161 of][]{Sneddon66}
\begin{equation}\label{eq:sneddon_ansatz}
    -2E^{(1)} + \sum_{n=1}^{\infty} \widehat{d}_n^{\,\,(1)} \cos(nz) = \cos\left(\frac{z}{2}\right)\int_{z}^{\gf\pi}\frac{h(t)}{\sqrt{\cos(z)-\cos(t)}}\,dt \qquad \text{for } |z| < \gf\pi.
\end{equation}
\fernando{H}ere $h(t)$ can be retrieved from \fernando{equation (5.4.60) in page 162 of \cite{Sneddon66}, which in our case simplifies to}
%
\begin{alignat}{2}\label{eq:sneddon_sol_h}
    h(t) &= \frac{2}{\pi}\frac{d}{dt}\int_{0}^{t}\frac{\sin\left(z/2\right)}{\sqrt{\cos(z)-\cos(t)}}\left(\int_{0}^{z}\left[-\frac{1}{4\pi}(1-\gamma_{Ma})\right]\,du\right)\,dz \nonumber\\
         &= -\frac{\sqrt{2}}{4\pi}(1-\gamma_{Ma})\tan\left(\frac{t}{2}\right),
\end{alignat}
where it is worth noting that the closed form of the integral
\begin{equation}\label{eq:insane_integral}
    \int_{0}^{t}\frac{z\sin\left(z/2\right)}{\sqrt{\cos(z)-\cos(t)}}\,dz = \sqrt{2}\,\pi\ln(\sec\left(t/2\right))
\end{equation}
has been used in the derivation above, obtained from \cite{sbragaglia07}.

The desired slip velocity for $0 \leq z < \gf\pi$ \fernando{can now be} retrieved \fernando{at leading order introducing the expansions \eqref{eq:alpha_beta_for_k_inf} and \eqref{eq:E_dn_for_k_inf} into \eqref{eq:uIallXAppdx} and using} equation (\ref{eq:sneddon_ansatz})
\begin{alignat}{3}
    u_{I}(z) &= L\left( 2E^{(1)} - \sum_{n=1}^{\infty} \widehat{d}_n^{\,\,(1)} \cos(nz) \right)       &&+ O(L^2) \nonumber\\
             &=-L\cos\left(\frac{z}{2}\right)\int_{z}^{\gf\pi}\frac{h(t)}{\sqrt{\cos(z)-\cos(t)}}\,dt\, &&+ O(L^2).
\end{alignat}
Making use of \fernando{\eqref{eq:sneddon_sol_h}}, integrating and undoing the change of variable we obtain the velocity profile
\begin{equation} \label{eq:uI_for_k_inf_Sbrag}
    u_{I}(x) = 
    \begin{cases}\displaystyle
    (1-\gamma_{Ma})\frac{L}{2\pi}\arccosh\left(\cfrac{\cos\left(\pi{x}/L\right)}{\cos\left(\pi\gf/2\right)}\right) + O(L^2) & \text{for } |x|\leq g/2,\\
    0                                                                                                                       & \text{for } g/2 \leq |x|\leq L/2.
    \end{cases}
\end{equation}
\fernando{The above formula \eqref{eq:uI_for_k_inf_Sbrag}}, after setting $\gamma_{Ma}=0$ and a change in the variables normalization, is \fernando{at leading order} exactly half of the slip velocity obtained by \cite{sbragaglia07} for a configuration with longitudinal no-shear infinite gaps in a semi-infinite domain. This result is consistent with the analysis of \cite{Asmolov2012}, who conclude that the slip velocity profile in such a configuration should be larger than that of the equivalent transverse case by exactly a factor of two.

\fernando{From \eqref{eq:uI_for_k_inf_Sbrag}, we can finally obtain the normalized slip velocity at mid-gap at leading order}
\begin{equation} \label{eq:uIcenter_for_k_inf_Sbrag}
    \cfrac{u_{Ic}}{2(1-\gamma_{Ma})} \simeq \frac{g}{4\pi\gf}\arccosh\left(\sec\left(\cfrac{\pi\gf}{2}\right)\right),
\end{equation}
where we have substituted $L=g/\gf$.

From \eqref{eq:uIcenter_for_k_inf_Sbrag} we can \fernando{corroborate the validity of} the linear scaling $u_I\sim{g}$ for $g\lesssim\nobreak{1}$. Indeed, the asymptote \eqref{eq:uIcenter_for_k_inf_Sbrag} is plotted for $\gf=0.99$ in \fig~\ref{fig:uIc_with_g_and_phi}($a$), showing good agreement with the numerically computed slip velocity.

Consider now the limit $\gf\rightarrow{0}$ within the regime of small gap length $g\rightarrow{0}$ investigated in this subsection. Then \eqref{eq:uIcenter_for_k_inf_Sbrag}  yields, to leading order in $g$,
\begin{equation}\label{eq:uIcenter_for_k_0_limit_phi_0}
    \frac{u_{Ic}}{2(1-\gamma_{Ma})} \sim \cfrac{g}{8} \qquad \text{for }\gf\rightarrow 0.
\end{equation}
This is congruent with the linear asymptote for small $g$ followed by the values calculated numerically, which is also shown in \fig~\ref{fig:uIc_with_g_and_phi}($a$).

\fernando{The coefficient $E$, which appears in the expressions for the effective slip length \eqref{eq:globaleffsliplength} and drag reduction \eqref{eq:DR}, can also be obtained in this limit from \eqref{eq:dual_series_k_inf} following \cite{Sneddon66}. We make use of \eqref{eq:sneddon_sol_h} to arrive at}

\begin{alignat}{2}\label{eq:E_for_k_inf}
    E = E^{(1)}L + O(L^2) &= -\cfrac{\sqrt{2}\,L}{4}\int_{0}^{\pi\gf}h(t)dt + O(L^2)  \nonumber\\
                          &= (1-\gamma_{Ma})\cfrac{L}{4\pi}\ln\left(\sec\left(\cfrac{\pi\gf}{2}\right)\right) + O(L^2)  
\end{alignat}

\section{Application of our model to experimental studies in the literature showing reduced slip}\label{apdx:discussion-PNAS-Song}

\julien{We study the experimental results of \cite{peaudecerf17} and \cite{Song2018-uw} to analyse with our theoretical model how surfactant  affected their SHS performance. The slip velocities extrapolated from the measurements of \cite{peaudecerf17}  on the interface ($z=0$) at mid-gap ($y=0$) are: $u_I\approx \SI{4e-3}{}\pm \SI{4e-3}{}$ for 2 mm long lanes (see their figure 3D), which is practically negligible; and $u_I\approx \SI{5e-2}{}\pm \SI{9e-3}{}$ for 30 mm long lanes (figure 3E), which is significantly reduced compared with the  theoretical (surfactant-free) prediction.  (Note that we have non-dimensionalised these velocities using the characteristic velocity $U$, following the convention used in the present study.) 
Similarly, \cite{Song2018-uw} report: $u_I\approx \SI{8e-2}{}$ for 5 mm long lanes (see their figure 3\textit{b}), which is significantly reduced compared with the  theoretical (surfactant-free) prediction; and $u_I\approx \SI{8e-3}{}$ for 15 mm long lanes (figure 5), which is practically negligible.
}

\julien{
The main difficulty in applying our theoretical model, for instance to predict the reduced slip velocities measured experimentally by \cite{peaudecerf17} and \cite{Song2018-uw}, is that the surfactant properties and their concentrations are completely unknown in their experiments. 
Instead, we use our model to predict the concentration of surfactant, for three different possible surfactant types, which could lead to the measured $u_I$ reported in \cite{peaudecerf17} and \cite{Song2018-uw}. The three surfactants we choose are: a `strong' poorly soluble surfactant with properties described in \cite{peaudecerf17}, a `weak' highly soluble surfactant, namely Sodium Dodecyl Sulfate (SDS), and an `intermediate' type with similar weak properties as SDS but rendered almost insoluble in water by reducing its desorption coefficient to $\hat{\kappa}_d=\SI{1}{s^{-1}}$ (instead of $\hat{\kappa}_d=\SI{500}{s^{-1}}$ for  SDS in water). 
Surfactants have a large number of parameters  ($\hat{\kappa}_d$, $\hat{\kappa}_a$, $\hat{D}$, $\hat{D}_I$, $\hat{\Gamma}_m$, $A$, $n_\sigma$), in addition to their bulk background concentration $\hat{c}_0$, which are almost all used in our theoretical model (see (\ref{eq:scalinguIsurfactant}) and (\ref{eq:gammaMa}), which we use to compute $u_I$. Thus, by choosing only three different types of surfactants from the vast parameter space, the analysis in this section is primarily qualitative. The aim is  to show that our theoretical model  provides physically meaningful explanations regarding the impact of surfactant in  experimental studies showing reduced SHS performance such as those of \cite{peaudecerf17} and \cite{Song2018-uw}.}

\julien{Assuming a strong surfactant, our  model predicts that a bulk surfactant concentration $\hat{c}_0\sim \SI{e-13}{mM}$ can reduce $u_I$ in the same extent and under the same conditions as reported by \cite{peaudecerf17} for both short and long lanes; and $\hat{c}_0\sim \SI{e-15}{}$ to $\SI{e-14}{mM}$ for the experiments reported by \cite{Song2018-uw}. Assuming the weak SDS surfactant, our  model predicts  $\hat{c}_0\sim 1$ to $\SI{10}{mM}$ (\ie near the critical micellar concentration) to obtain the results of \cite{peaudecerf17} and $\hat{c}_0\sim 0.1$ to $\SI{3}{mM}$ for the experimental results of \cite{Song2018-uw}. Assuming an intermediate surfactant,  $\hat{c}_0\sim \SI{e-5}{}$ to $\SI{e-4}{mM}$ would lead to the  results of \cite{peaudecerf17}, and $\hat{c}_0\sim \SI{4e-7}{}$ to $\SI{e-5}{mM}$ for the results of \cite{Song2018-uw}. These theoretical predictions show that: (i) a very strong surfactant would require only minute traces,  unavoidable in normal environmental conditions, to strongly affect slip; (ii) whilst at the other extreme, a weak surfactant such as SDS would require a concentration of the order of the critical micellar concentration to lead to a no-slip or reduced slip condition. Then, an intermediate surfactant  would require  small concentration at or below typical environmental background concentration to lead to no-slip or reduced slip condition. 
}

\julien{Therefore, our theoretical model provides physically sensible predictions with regard to surfactant types and concentrations that may have contaminated the experiments of \cite{peaudecerf17} and \cite{Song2018-uw}. This is consistent with their conclusions.  We  note that our theoretical model assumes a two-dimensional channel geometry with one-dimensional interfaces, whereas the experiments are three-dimensional with two-dimensional (flat) or three-dimensional (curved) interfaces   bounded laterally by no-slip  walls. Hence, we expect our model to over-predict the interfacial slip velocity (for a given surfactant type and concentration) or over-predict the background surfactant concentration (for a given surfactant type and interfacial slip velocity). This means that even lower surfactant concentrations could have affected the experimental results  of \cite{peaudecerf17} and \cite{Song2018-uw}.
}

\end{document}